\documentclass[twocolumn,prb,aps,scriptaddress]{revtex4-2}
\usepackage[utf8]{inputenc}
\usepackage{color}
\usepackage{amsmath}
\usepackage{graphicx}
\usepackage{comment}
\usepackage{multirow}
\usepackage[bookmarks=false,
 breaklinks=false,pdfborder={0 0 1},backref=section,colorlinks=false]
 {hyperref}

\makeatother

\makeatother
\begin{document}

\title{Anomalous behavior of native point defects in C2-ordered antiferromagnet $\alpha$-MnO$_2$}

\author{Archana Sharma}
\affiliation{Department of Physics, Indian Institute of Technology Bombay, Mumbai 400076, India}
\author{Brahmananda Chakraborty}
\email{brahma@barc.gov.in}

\affiliation{High Pressure and Synchrotron Radiation Physics Division, Bhabha Atomic Research Centre, Trombay, Mumbai 400085, India}
\altaffiliation[Also at ]{Homi Bhabha National Institute, Mumbai 400085, India}
             
\begin{abstract}

$\alpha$-MnO$_2$ is an emerging material for electronic, optoelectronic, and energy applications, owing to its structural flexibility and defect-driven functionality. During synthesis of $\alpha$-MnO$_2$, native oxygen vacancies readily form and are typically compensated by foreign dopants. A thorough understanding of intrinsic defects is therefore essential for enabling controlled extrinsic doping and optimizing material performance. Using density functional approach, we investigate the structural, electronic, magnetic, and optical properties of the ground state C2-type antiferromagnetic $\alpha$-MnO$_2$ in the presence of native point defects, including interstitials, vacancies, and antisites. We compute their thermodynamic stability, incorporating electrostatic corrections to eliminate spurious long-range interactions. Mn interstitial (Mn$_\text{i}$) and Mn antisite O (Mn$_\text{O}$) introduce shallow donor levels, whereas O-vacancy (V$_\text{O}$) exhibit amphoteric behavior and act as compensating centers. The calculated defect formation energies reveal pronounced competition between donor- and acceptor-type native defects, leading to strong intrinsic defect compensation under both Mn-rich and O-rich growth conditions. The presence of defects significantly perturbs the electronic structure of $\alpha$-MnO$_2$, introducing spin-polarized midgap states while O interstitial (O$_\text{i}$) preserves the host antiferromagnetic order. Mn vacancy (V$_\text{Mn}$) remains ionized across the band gap and behaves as a shallow acceptor, suggesting its potential role under suitable non-equilibrium growth conditions, whereas O antisite Mn (O$_\text{Mn}$) forms deep acceptor levels. BSE@G$_0$$W_0$ calculations reveal a strongly anisotropic optical response in stoichiometric $\alpha$-MnO$_2$, while native point defects introduce pronounced sub-gap excitations and enhanced dielectric screening, with vacancies producing the largest effect.

\end{abstract}

\maketitle

\section{\label{sec:level1}Introduction}

Manganese oxides, with various polymorphs and stoichiometries, have attracted significant attention due to their natural abundance, environmental compatibility, and high chemical reactivity~\cite{franchini2007ground}. Various MnO$_2$ phases such as pyrolusite ($\beta$), ramsdellite (R), hollandite ($\alpha$), intergrowth ($\gamma$), spinel ($\lambda$), and layered ($\delta$) have been widely used in technologies including catalysis, batteries, supercapacitors, solid-state ionics, magnetoresistive devices, filtration, and sensors~\cite{kitchaev2016energetics,yang2021mno2,davoglio2018synthesis,xiankai2024novel,taranu2022alpha}. Among these, $\alpha$-MnO$_2$ has emerged as a particularly promising candidate due to its unique tunnel framework, low cost, and ease of synthesis~\cite{taranu2022alpha}.

$\alpha$-MnO$_2$ crystallizes in a tetragonal structure (space group \textit{I4/m}), comprising double chains of edge-sharing MnO$_6$ octahedra \cite{bystrom1950crystal}. The Mn$^{4+}$ ions adopt a spin-polarized 3$d^3$ configuration, while O$^{2-}$ ions remain spin-unpolarized in 2$p^6$ states~\cite{kitchaev2016energetics}. Corner-sharing connectivity among octahedra forms 2$\times$2 tunnels (0.46\,nm $\times$ 0.46\,nm) that can accommodate cations and molecules~\cite{kaltak2017charge}. This structural flexibility enables tuning of the electrochemical and physical properties of $\alpha$-MnO$_2$ via ion intercalation or doping~\cite{young2015charge,luo2010tuning,lubke2018transition,lambert2017understanding}. The material is known to exhibit antiferromagnetic (AFM) ordering of the C2-type below the N\'eel temperature ($T_\text{N} = 24.5$\,K), with AFM coupling between corner-sharing Mn atoms and weak ferromagnetic interactions between edge-sharing octahedra~\cite{yamamoto1974single}.

Despite its technological relevance, stoichiometric $\alpha$-MnO$_2$ is difficult to synthesize, as the tunnels tend to trap stabilizing cations (e.g., K$^+$, Ba$^{2+}$) or water molecules following oxygen vacancy formation~\cite{yuan2016influence,gao2008microstructures}. While these dopants help stabilize the framework, they also alter the Mn oxidation states and induce Jahn–Teller distortions, lowering the symmetry to monoclinic (\textit{I2/m})~\cite{yuan2016influence,hou2013tuning,li2005synthesis}. Consequently, experimentally synthesized $\alpha$-MnO$_2$ often appears in cryptomelane or hollandite forms, depending on the tunnel-occupying species~\cite{cockayne2012first,gao2008microstructures}. These structural and chemical modifications improve conductivity and facilitate catalytic properties in $\alpha$-MnO$_2$~\cite{kaltak2017charge,yuan2016influence,hou2013tuning}. For example, Fe-doped $\alpha$-MnO$_2$ has demonstrated enhanced catalytic activity in oxygen evolution and industrial-scale photocatalysis~\cite{said2018photo,mathur2019one}.

While the role of extrinsic doping has been explored extensively~\cite{young2015charge,wang2020high,li2022highly}, the nature, energetics, and electronic signatures of intrinsic point defects in stoichiometric $\alpha$-MnO$_2$ remain largely unexplored. This is a critical knowledge gap, as native point defects such as vacancies, interstitials, and antisites can introduce localized states in the band gap, act as charge trapping centers, and significantly influence the electronic structure, magnetic ordering, and transport behavior. Interestingly, the ease of oxygen vacancy formation due to the presence of exposed MnO$_6$ octahedra suggests a natural tendency for charge imbalance in $\alpha$-MnO$_2$, often compensated by tunnel cation occupancy ~\cite{yuan2016influence,gao2008microstructures,wang2020high}. Therefore, understanding the thermodynamic and electronic behavior of native defects is crucial not only for evaluating the intrinsic conductivity and defect tolerance of $\alpha$-MnO$_2$, but also for guiding extrinsic doping, as native defects often dominate compensation behavior and govern dopant incorporation and effectiveness.

However, the electronic structure of stoichiometric $\alpha$-MnO$_2$ remains poorly characterized. Experimental measurements of the band gap are inconsistent, with reported values ranging from 1.16 to 2.23 eV depending on sample morphology and dimensional confinement~\cite{gao2008microstructures,gangwar2021structural,liu2018near,salari2020facile,sakai2005photocurrent}. These discrepancies arise partly from the presence of dopants and the lack of measurements below the N\'eel temperature, implying that the true AFM ground state is rarely probed experimentally~\cite{yamamoto1974single,trimarchi2018polymorphous}. Since the electronic structure of transition metal oxides is sensitive to magnetic ordering, this complicates comparison between theory and experiment. To better understand its inherent semiconducting behavior, we compute more accurate electronic band structure and band gap using G$_0$$W_0$ approximation.

In this work, we present a comprehensive first-principles study of native point defects in stoichiometric $\alpha$-MnO$_2$, which, to the best of our knowledge, has not been reported previously. Using density functional theory calculations, we systematically investigate the formation energies and charge transition levels of oxygen vacancy (V$_\text{O}$), oxygen interstitial (O$_\text{i}$), manganese interstitial (Mn$_\text{i}$), manganese vacancy (V$_\text{Mn}$), manganese antisite on oxygen site (Mn$_\text{O}$), and oxygen antisite on manganese site (O$_\text{Mn}$), including finite-size corrections. We evaluate the thermodynamic stability of these defects under both Mn-rich and O-rich conditions and analyze their influence on the electronic structure, optical properties and dielectric response. Our results provide crucial insights into the intrinsic defect physics and optical activity of $\alpha$-MnO$_2$, laying the foundation for defect engineering and controlled doping strategies to tailor its electronic, opto-electronic, and catalytic performance for device applications.

\section{\label{sec:level1}Computational Methods}

First-principles calculations are performed in the density functional theory (DFT) framework with spin polarization, implemented in the Vienna \textit{ab initio} simulation package (VASP) code~\cite{vasp1}. The Perdew-Burke-Ernzerhof (PBE) exchange-correlation functional is treated within the generalized gradient approximation (GGA)~\cite{pbe}. All calculations are performed with collinear spin configurations, with the magnetic moments initialized to converge to the ground state C2-type AFM ordering. The Brillouin zone is sampled with a $\Gamma$-centered 4$\times$4$\times$4 and 4$\times$4$\times$8 \textit{k}-mesh of the Monkhorst-Pack scheme~\cite{monkhorst} for primitive and conventional unit cell optimization, respectively. The electronic wave functions are expanded in the projector augmented wave (PAW) basis set~\cite{paw}, employing a cutoff energy of 450 eV. Semicore potentials of Mn (3$s^2$3$p^6$4$s^2$3$d^5$) and O (2$s^2$2$p^4$) are included due to the importance of core-valence exchange in $d$-electron systems~\cite{engel2009relevance}. The Gaussian smearing method is used with a smearing parameter of 0.05 eV. 

To simulate isolated defects, we choose a 1$\times$1$\times$3 supercell of $\alpha$-MnO$_2$ to reduce the interaction between periodic defect images, particularly in the c-lattice direction, and the Brillouin zone is sampled using a 4$\times$4$\times$4 \textit{k}-point grid. The atoms in the lattice containing defects are allowed to relax until the maximum force on an atom is 5 meV/\AA~and the energy tolerance of $10^{-7}$ eV for electronic iterations is reached. The presence of strongly correlated 3$d$ electrons in Mn requires the inclusion of inter-atomic interactions through the Hubbard potential ($U$). Thus, all structures are relaxed using the rotationally invariant PBE+$U$ method by Dudarev~\cite{dudarev}, followed by electronic structure calculations. The on-site Coulomb interaction for Mn $3d$ states is determined through a systematic assessment over the range $U$ = 2-5 eV. For each value of $U$, the structural parameters, Mn magnetic moments, electronic band gap, and the total-energy difference between the C2 antiferromagnetic (AFM) and ferromagnetic (FM) configurations are evaluated. The experimentally observed C2 AFM ordering is found to be energetically stable for $U \le$ 3.8 eV, whereas larger $U$ values favor a FM ground state. The band gap increases with $U$ up to 3.8 eV and decreases thereafter. The optimized lattice parameters and Mn magnetic moment at $U$ = 3.8 eV are in good agreement with available experimental data \cite{thackeray1997manganese,li2007one}. Based on the correct magnetic ground state and its suitability as a starting point for subsequent G$_0$W$_0$ calculations, $U$ = 3.8 eV is adopted throughout this work. The detailed benchmarking results are provided in Fig. S1 of the Supplemental Material (SM) \cite{SM}.

\begin{figure*}[ht]
\centering
\includegraphics[width=0.7\linewidth]{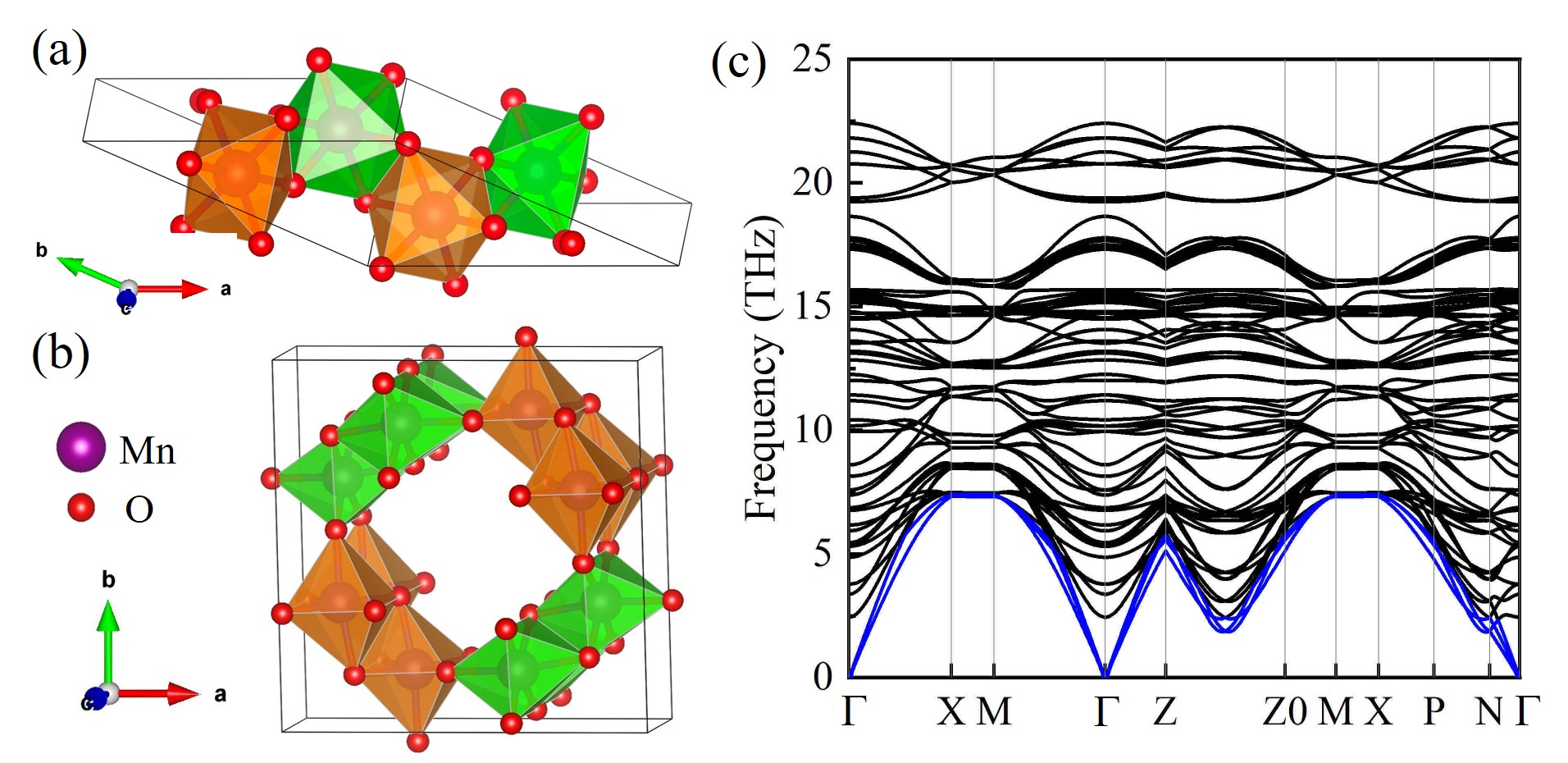}
\caption{Crystal structures of $\alpha$-MnO$_2$ in C2-AFM configuration: (a) primitive unit cell, and (b) conventional tetragonal structure. Spin up and spin down alignments are represented by green and orange octahedra. (c) Calculated phonon dispersion spectra of the conventional lattice along the high-symmetry points. The three acoustic modes near the $\Gamma$-point are shown in blue.}
\label{fig1}
\end{figure*}

We have also calculated quasiparticle band gaps using the single-shot G$_0$$W_0$ approximation~\cite{gw1,gw2} on the PBE+$U$ wavefunctions. The response function is calculated using a \( \Gamma \)-centered 2$\times$2$\times$4 \textit{q}-point mesh. We have utilized 768 valence and conduction bands and 100 frequency points, ensuring that the quasiparticle band gap (QP) is well converged. 
The band energies are interpolated through maximally localized Wannier functions (MLWFs), as implemented in the WANNIER90 code~\cite{Ref83}. The initial projections for Mn and O atoms are chosen to be (s, d) and (s, p) orbitals, respectively. The \textit{k}-point path is generated through the SeeK-path code \cite{hinuma2017band}. The oxidation state (OS) of the Mn ion is determined using the projection-based method by Sit \textit{et al.}~\cite{sit2011simple}, which has been effective in determining the OS of transition metal elements in various complex materials~\cite{ku2019oxidation,timrov2022accurate}. The phonon dispersion curves are calculated using linear response density functional perturbation theory (DFPT)~\cite{gonze1997dynamical} as implemented in the PHONOPY code~\cite{togo2023implementation}.

The defects analyzed in this study include interstitials (Mn$_\text{i}$, O$_\text{i}$), vacancies (V$_\text{Mn}$, V$_\text{O}$), and antisites (Mn$_\text{O}$, O$_\text{Mn}$). We calculate the formation energies of these defects using a supercell approach, according to
\begin{equation}
    E_{f}(D^q) = E_\mathrm{tot}(D^q) - E_\mathrm{tot}(\text{bulk}) - \sum_j n_j \mu_j + qE_F + E_\mathrm{corr},
    \label{eq1}
\end{equation}
where $E_\mathrm{tot}(\text{bulk})$ is the total energy per supercell of stoichiometric $\alpha$-MnO$_2$, $E_\mathrm{tot}(D^q)$ is the total energy with a defect of charge $q$, $n_j$ is the number of atoms of element $j$ (Mn or O) added to or removed from a defect-free cell, to its respective reservoir with chemical potential, $\mu_j$ to form the defect cell. $E_F$ is the Fermi energy referenced to the valence band maximum (VBM), and $E_\mathrm{corr}$ is the electrostatic correction term, discussed in Sec. \ref{sec:level2}.

While formation energies help predict defect concentrations and stabilities, the position of charge transition levels (CTLs)~\cite{freysoldt2014first} is important to identify whether the defects are shallow or deep and whether they act as donors or acceptors. The CTLs are calculated by:
\begin{equation}
    \varepsilon(q/q') = \frac{E^f(D^q) - E^f(D^{q'})}{q' - q},
\end{equation}
indicating the Fermi level at which the stable charge state transitions from $q$ to $q'$.

The excitonic optical properties are computed by solving the Bethe–Salpeter equation (BSE) \cite{rohlfing2000electron} on top of single-shot G$_0$W$_0$ quasiparticle energies and wave functions. The BSE kernel includes the screened direct electron–hole attraction and the unscreened exchange interaction. A total of 180 valence and 84 conduction bands are included in constructing the BSE Hamiltonian to achieve sufficient convergence of the low-energy optical response. The macroscopic dielectric function is evaluated within the Tamm–Dancoff approximation, which has been shown to provide reliable optical spectra for transition-metal oxides.

\section{Results and Discussion}

\subsection{Bulk properties of $\alpha$-MnO$_2$ crystal}

The lattice parameters for the C2-AFM magnetic configuration of $\alpha$-MnO$_2$ relaxed using the PBE+$U$ functional (for $U$ = 3.8 eV) are calculated as a = 9.83 \AA~ and c = 2.91 \AA~, which are in good agreement with the experimental results \cite{thackeray1997manganese}. The local magnetic moment on each Mn is 3.14 $\mu_B$, which is very close to the one reported by Li et al. ~\cite{li2007one}. The oxidation state of each Mn ion is Mn(IV), corresponding to 3$d^3$ configuration. This shows a uniform octahedral crystal field splitting between the 3$d^3$ states of Mn ions into threefold degenerate, occupied $t_{2g}$ states and the double degenerate, unoccupied $e_g$ states. Each Mn atom is coordinated to six O atoms, forming an octahedron as shown in Fig. \ref{fig1}(a)and \ref{fig1}(b). The calculated Mn-O bond length for each of the six bonds is 1.93 \AA~, which closely matches the experimental values~\cite{kijima2004crystal}. The inclusion of spin-orbit coupling (SOC) has a negligible effect on the band structure, as illustrated in Fig. S2. Consequently, SOC effects are omitted in the subsequent calculations for computational efficiency without compromising accuracy. Fig. \ref{fig1}(c) presents the calculated phonon band structure of $\alpha$-MnO$_2$ with the conventional cell. The absence of imaginary frequencies in the phonon dispersion confirms the dynamic stability of the structure at 0 K. For $\alpha$-MnO$_2$, the dielectric response is anisotropic, and the dielectric tensor $\boldsymbol{\epsilon}_\infty$ is computed using DFPT~\cite{souza2002first}, yielding values of 8.25 and 9.49 along the $a$ and $c$ directions, respectively, consistent with the underlying crystal anisotropy.

Several experimental and theoretical studies have reported a wide range of band gap values for $\alpha$-MnO$_2$, largely influenced by sample morphology, measurement techniques, and magnetic ordering. For instance, the optical band gap estimated from UV-visible absorption spectra for $\alpha$-MnO$_2$ nanorods is reported to be 1.53 eV~\cite{gangwar2021structural}, which closely matches the 1.65 eV value observed by Salari et al.~~\cite{salari2020facile}. A lower optical gap of 1.16 eV has also been reported~~\cite{liu2018near}, aligning well with the value of 1.18 eV obtained from vertex-corrected GW calculations~~\cite{abdallah2025electronic}. Sakai et al. found that ultrathin $\alpha$-MnO$_2$ nanosheets (0.5 nm thick) exhibit a band gap of 2.23 eV ~\cite{sakai2005photocurrent}, a shift attributed to quantum confinement effects. Similarly, Gao et al. reported a band gap of 1.32 eV in $\alpha$-MnO$_2$ nanofibers with diameters of 20-60 nm and lengths of 1-6 $\mu_m$ \cite{gao2008microstructures}. Using the PBE+$U$ functional, band gaps of 1.3 eV~~\cite{cockayne2012first} and 1.28 eV~~\cite{chepkoech2018first} are obtained with U = 1.6 eV and U = 2.4 eV, respectively. Crespo et al. \cite{crespo2013electronic} obtained a larger value of 2.2 eV using the HSE06 hybrid functional, although their calculations predicted a ferromagnetic ground state. Vertex-corrected self-consistent GW approaches yield band gaps of 1.67-1.94 eV ~\cite{abdallah2025electronic}.

\begin{figure}[ht]
\centering
\includegraphics[width=1.0\linewidth]{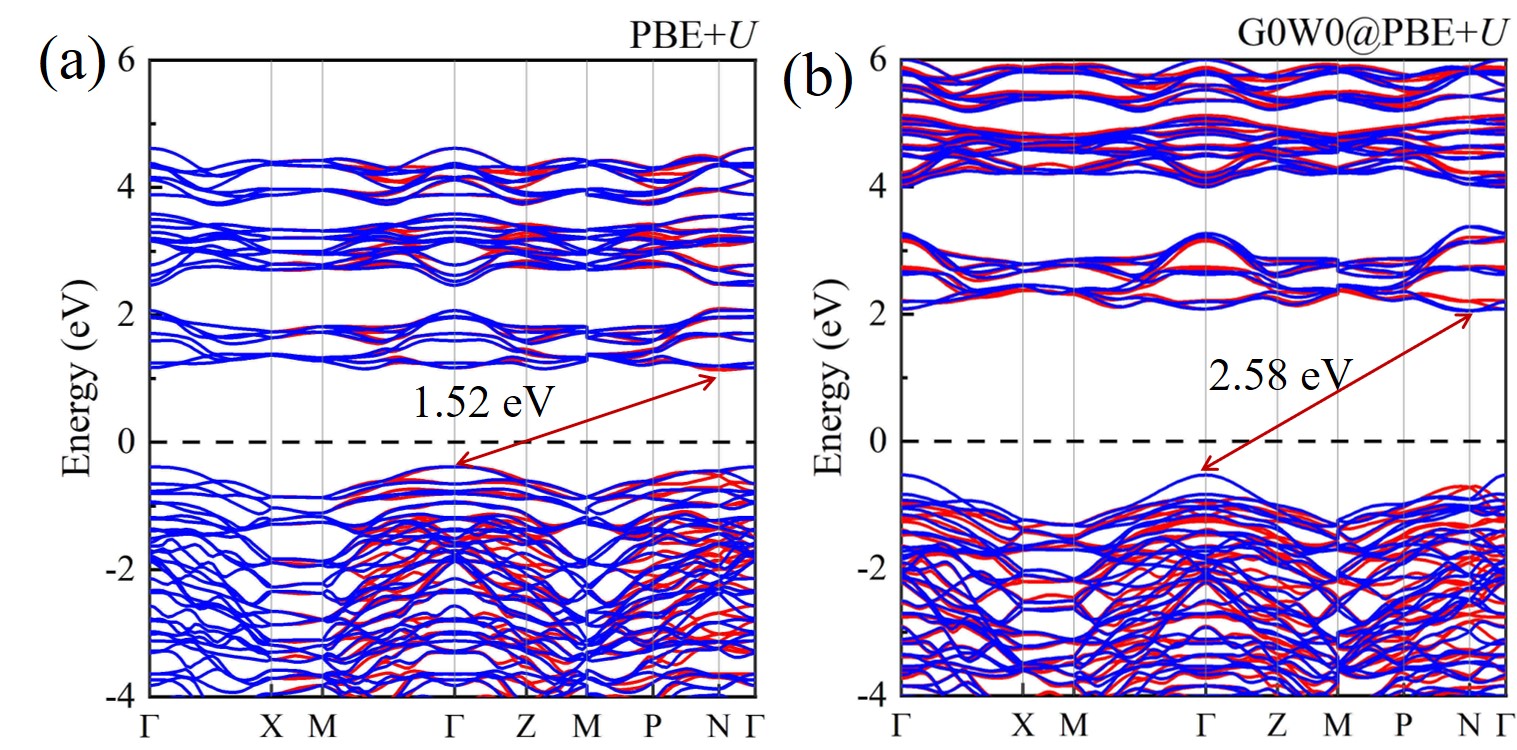}
\caption{(a),(b) Spin-polarized electronic band structures of conventional cell of $\alpha$-MnO$_2$ using DFT+$U$ and DFT+G$_0$$W_0$ level of theory. High symmetry points Z\textbar Z0 and M\textbar X are denoted by Z and M, for brevity. Red(blue) color denotes spin-up(spin-down) bands. The Fermi level is scaled to 0 eV.}
\label{fig2}
\end{figure}

In our calculations, we find an indirect band gap of 1.52 eV using the PBE+$U$ method [cf. Fig. \ref{fig2}(a)], which is in close agreement with previous theoretical and experimental results. The G$_0$$W_0$@PBE+$U$ calculated band gap increases to 2.58 eV as shown in Fig. \ref{fig2}(b). Given the overall agreement of PBE+$U$ with experimental estimates and its computational efficiency, we adopt it for subsequent discussions. All the computed parameters discussed in this section are presented in Table~\ref{table1}.

\begin{table}[ht]
\caption{\label{table1}
Lattice constants ($a$, $c$), local magnetic moment on Mn ($\mu_\text{Mn}$), band gap ($E_g$) from PBE+$U$ and G$W_0$ methods, and macroscopic dielectric constant ($\varepsilon_\infty$) along a-axis ($\varepsilon_a$) and c-axis ($\varepsilon_c$) for stoichiometric $\alpha$-MnO$_2$. Experimental and theoretical references from literature are included for comparison.}
\begin{ruledtabular}
\begin{tabular}{lcccc}
 & $a$, $c$ (\AA) & $\mu_\text{Mn}$ ($\mu_B$) & $E_g$ (eV) & $\varepsilon_\infty$ \\
\hline
This work & 9.83, 2.91 & 3.14 & \begin{tabular}{@{}c@{}}1.52 (PBE+$U$)\\2.58 (G$_0$$W_0$)\end{tabular} & \begin{tabular}{@{}c@{}}8.25 ($\varepsilon_a$)\\9.49 ($\varepsilon_c$)\end{tabular}\\
Exp. & 9.75, 2.86\footnotemark[1] & 3.94-4.04\footnotemark[2] & 1.16-2.23\footnotemark[3] & -\\
Theory & Ref.~\cite{crespo2013electronic,cockayne2012first} & \cite{chepkoech2018first,mahajan2022pivotal} & 1.18-2.20\footnotemark[4] & Ref.~\cite{abdallah2025electronic}\\
\end{tabular}
\end{ruledtabular}
\footnotetext[1]{Ref.~\cite{thackeray1997manganese}}
\footnotetext[2]{Ref.~\cite{li2007one}}
\footnotetext[3]{Ref.~\cite{gao2008microstructures,gangwar2021structural,liu2018near,salari2020facile,sakai2005photocurrent}}
\footnotetext[4]{Ref.~\cite{cockayne2012first,abdallah2025electronic,chepkoech2018first,crespo2013electronic}}
\end{table}

\subsection{Electronic structure of isolated defects}

\begin{figure}[ht]
    \centering
    \includegraphics[width=1\linewidth]{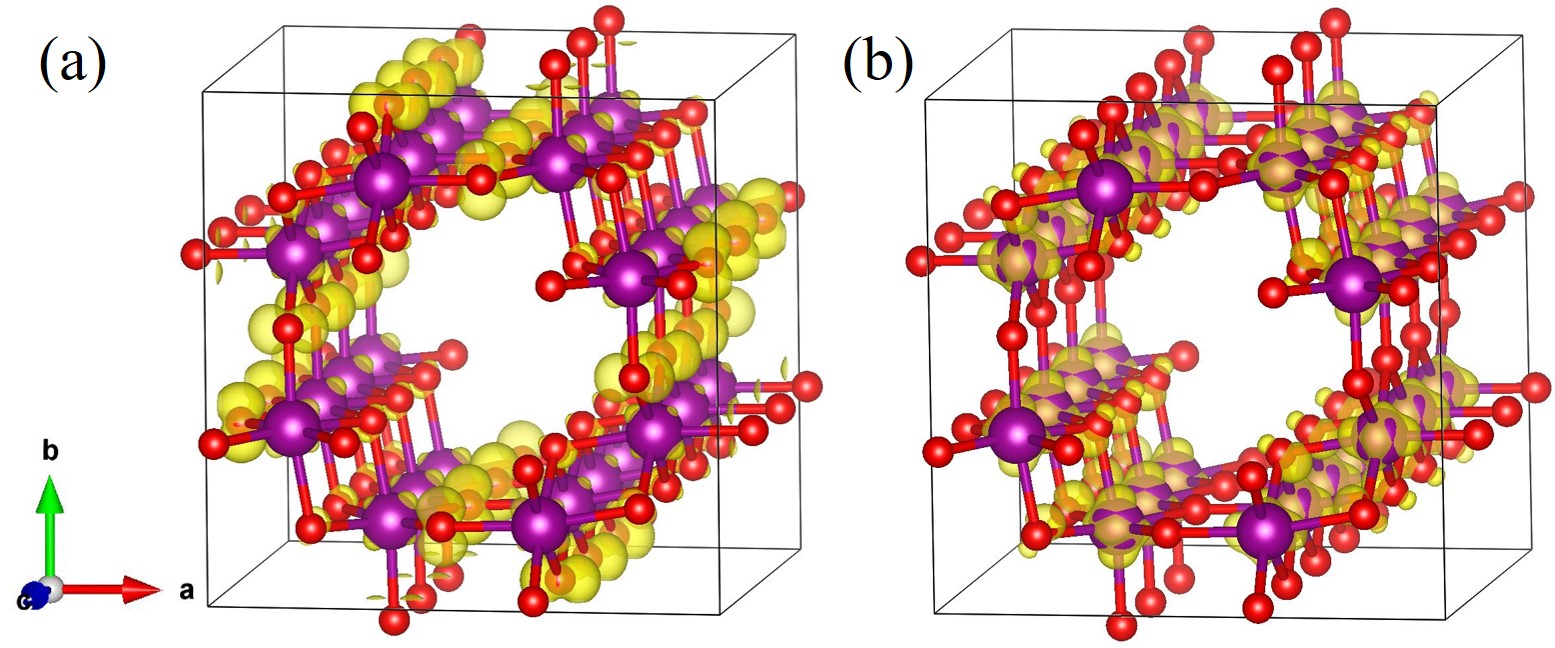}
    \caption{(a), (b) Band-decomposed spin charge densities of the VBM and the CBM  for the \(1 \times 1 \times 3\) $\alpha$-MnO$_2$ supercell. Isosurfaces are plotted with a value of \(2.6 \times 10^{-3}\,e\,\text{\AA}^{-3}\).}
    \label{fig3}
\end{figure}

To investigate the electronic properties of defective $\alpha$-MnO$_2$, we begin with an analysis of the stoichiometric material using a \(1 \times 1 \times 3\) supercell. Figs.~\ref{fig3}(a) and \ref{fig3}(b) show the band-decomposed charge densities corresponding to the valence band maximum (VBM) and conduction band minimum (CBM), respectively. The VBM is predominantly composed of O-2\textit{p} states, while the CBM exhibits hybridized contributions from both Mn-3\textit{d} and O-2\textit{p} orbitals. The computed electronic band structure, presented in Fig. S3(a), reveals significant band folding effects due to the enlarged unit cell, resulting in an increased number of bands within the Brillouin zone. The orbital-projected density of states (PDOS), shown in Fig.~\ref{fig5}(a), confirms the dominance of O-2\textit{p} states in the valence band, with moderate Mn-3\textit{d} participation, while the conduction band consists primarily of Mn-3\textit{d} and O-2\textit{p} hybridized states. Although experimental data on the band edges of stoichiometric $\alpha$-MnO$_2$ is currently unavailable, these results provide a crucial baseline for identifying defect-induced states and evaluating the semiconducting character of defective structures.

\begin{figure*}[ht]
    \centering
    \includegraphics[width=1\linewidth]{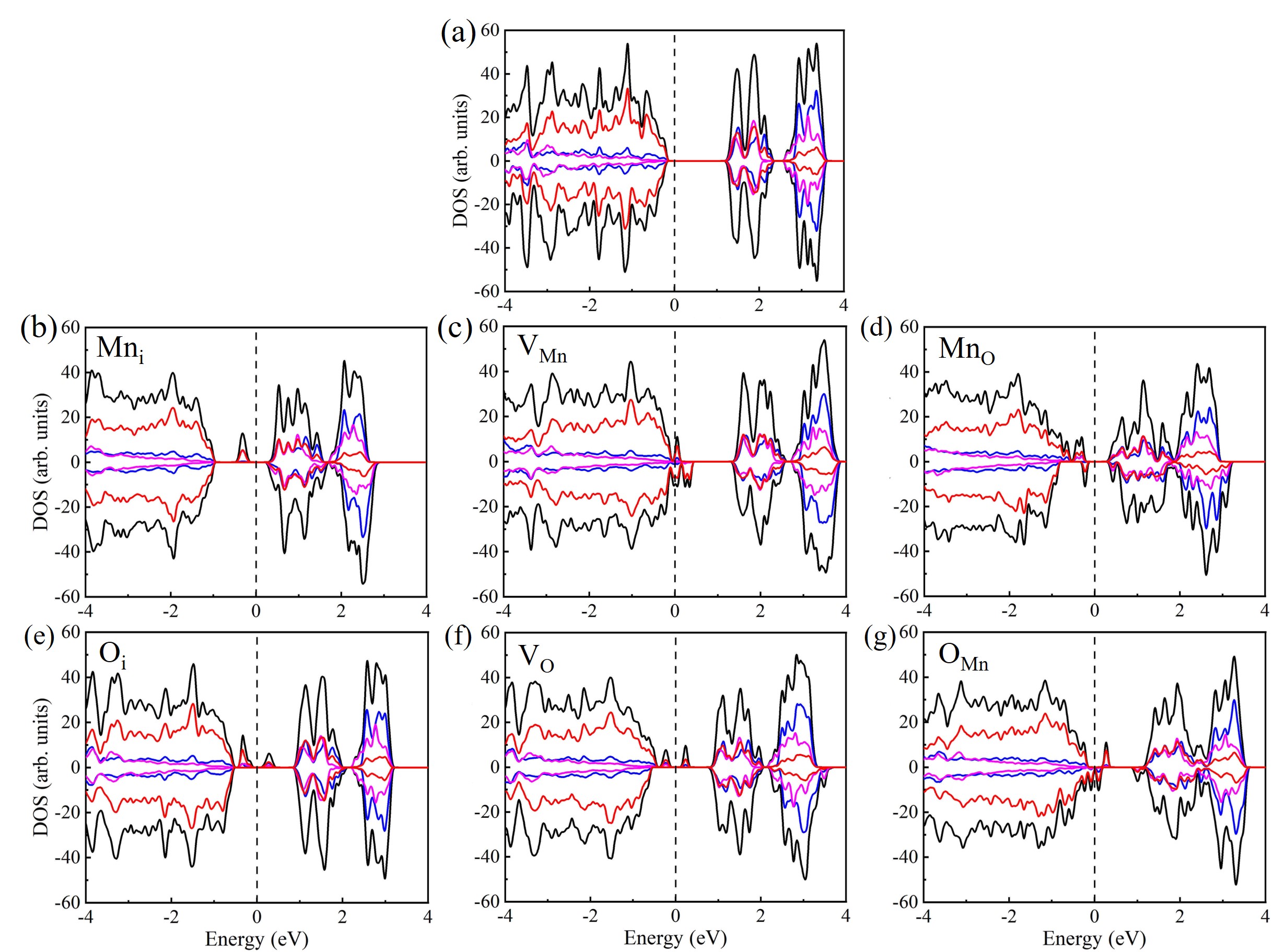}
    \caption{Local density of states (DOS) projected onto Mn-\(t_{2g}\) (blue), Mn-\(e_g\) (magenta), and O-2\(p\) (red) orbitals for (a) \(1 \times 1 \times 3\) $\alpha$-MnO$_2$ supercell without defects, (b)-(g) with intrinsic point defects.
    Total DOS is shown in solid black line. The Fermi level is scaled to 0 eV.}
    \label{fig4}
\end{figure*}

\begin{figure*}[ht]
    \centering
    \includegraphics[width=1\linewidth]{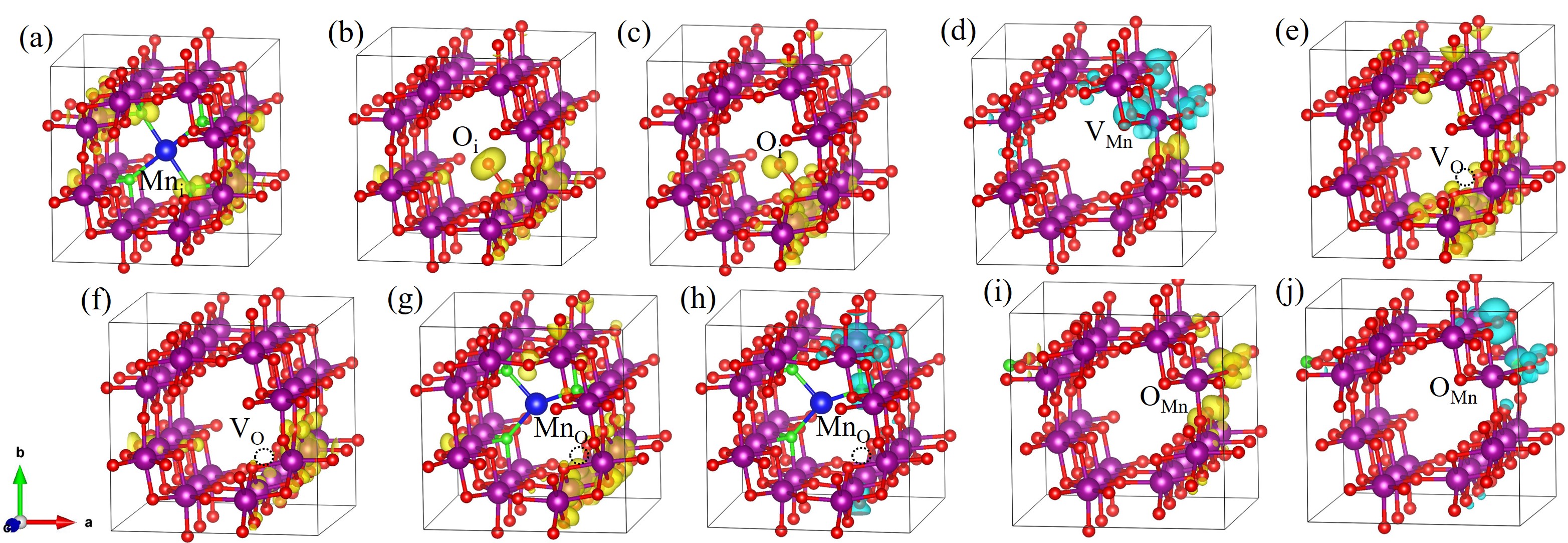}
    \caption{Defect wave functions of \(1 \times 1 \times 3\) $\alpha$-MnO$_2$ supercell with (a) a Mn interstitial (Mn\(_\text{i}\)); (b), (c) an O interstitial (O\(_\text{i}\)); (d) single Mn vacancy (V\(_\text{Mn}\)); (e), (f) single O vacancy (V\(_\text{O}\)); (g), (h) Mn antisite O (Mn\(_\text{O}\)); (i), (j) O antisite Mn (O\(_\text{Mn}\)). Isosurfaces are plotted with charge density values (in \(10^{-3}\,e\,\text{\AA}^{-3}\)) of 5.4, 7.6, 6.6, 4.5, 9.1, and 1.7, respectively. Spin-up and spin-down densities are represented by yellow and cyan isosurfaces. The Mn interstitial atom is shown in blue, and neighboring O atoms in green. Dashed black circle denotes an O-vacancy.}
    \label{fig5}
\end{figure*}

We now examine the effect of intrinsic point defects, including interstitials, vacancies, and antisite defects, on the electronic structure of $\alpha$-MnO$_2$. The representative defect configurations are shown in Fig. S4. The corresponding projected density of states (PDOS) and the electronic band structures are presented in Fig.~\ref{fig4}(b)-\ref{fig4}(g) and Fig. S3(b)-S3(g), respectively. The charge density wave functions of the defect-induced states are shown in Fig.~\ref{fig5}.

\subsubsection{Interstitials}

Mn interstitial (Mn\(_\text{i}\)) incorporated at the center of the tunnel forms two symmetric bonds with nearby lattice O atoms residing in opposite octahedra [cf. Fig.~S4(a)]. The Mn\(_\text{i}\)–O bond lengths along the diagonals are 2.21~\AA{} and 2.39~\AA{}, respectively, with concomitant changes in adjacent O–Mn lattice distances (2.25~\AA{} and 2.01~\AA). The PDOS for the neutral Mn\(_\text{i}\) defect shows a spin-majority doublet defect state near the VBM, originating from hybridized Mn-\textit{e}\(_g\) and O-2\textit{p} orbitals [cf. [Fig.~\ref{fig4}(a)] and S3(b)]. The corresponding charge density reveals defect-state localization on the adjacent lattice O atoms and nearby lattice Mn ions in opposite octahedra, resulting in reduction of these Mn(IV) to Mn(III) [cf. Fig.~\ref{fig5}(a)]. Each reduced Mn contributes 1~$\mu_B$, while the Mn\(_\text{i}\) retains its oxidation state of Mn(II), contributing 5~$\mu_B$. The net magnetic moment of the system increases to 7~$\mu_B$. Notably, the Mn\(_\text{i}\), characterized by a filled \(t_{2g}^3e_g^2\) configuration, remains spectroscopically inert with respect to charge addition or removal.

Interstitial O atom (O\(_\text{i}\)) is investigated at two distinct sites: within the tunnel center and adjacent to a lattice O atom. The latter configuration is energetically more favorable by 1.89~eV/defect, forming a dimer with a neighboring O atom [cf. Fig.~S4(b)]. The resulting O\(_\text{i}\)–O bond length is 1.30~\AA, intermediate between those in O\(_2\) (1.21~\AA) and peroxide (1.47~\AA). The PDOS and spin-polarized band structure display doubly occupied and singly unoccupied antibonding pp* states near the Fermi level, while bonding pp orbitals lie deeper in the valence band [cf. Figs.~\ref{fig4}(e) and Fig.~S4(e)]. The presence of electrons in the antibonding orbitals are responsible for enlarged O\(_\text{i}\)-O bond length as compared to free O\(_2\). The corresponding wavefunctions [cf. Figs.~\ref{fig5}(b), \ref{fig5}(c)] indicate that the charge localization occurs on both O\(_\text{i}\)-O dimer and surrounding three Mn ions. One of the Mn ions is reduced to Mn(III) (due to electron localization) [cf. Fig. \ref{fig5}(b)], while holes localize on other neighboring Mn sites [cf. Fig. \ref{fig5}(c)]. Furthermore, despite localized electron charges on both O\(_\text{i}\)-O dimer and the reduced Mn ion, the net magnetic moment remains zero.

\subsubsection{Vacancies}

Mn vacancy (V\(_\text{Mn}\)) disrupts six Mn–O bonds, effectively removing three electrons and leaving six oxygen dangling bonds [cf. Fig.~S4(c)]. These O atoms relax slightly outwards by $\sim$0.08~\AA due to the electron-deficient region at the vacant site. The electronic structure of neutral V\(_\text{Mn}\) shows defect-induced unoccupied states arising primarily from O 2$p$ orbitals, which appear near the the Fermi level [cf. Fig.~\ref{fig4}(c)]. Additionally, a few occupied defect states are observed close to the VBM, although they are unlikely to contribute significantly to conduction. Interestingly, one of the defect bands near the VBM touches the Fermi level, as shown more distinctly in Fig. S3(c). The presence of this partially filled, oxygen-derived state at the Fermi level indicates the formation of delocalized hole states. This is corroborated by the defect charge density, which shows hole delocalization over the six uncoordinated oxygen atoms surrounding the Mn vacancy [cf. Fig.~\ref{fig5}(d)], resulting in a net magnetic moment of 5 $\mu_B$. These findings highlight the role of V\(_\text{Mn}\) in modulating both the electronic and magnetic response of $\alpha$-MnO$_2$.

\begin{figure}[ht]
    \centering
    \includegraphics[width=0.9\linewidth]{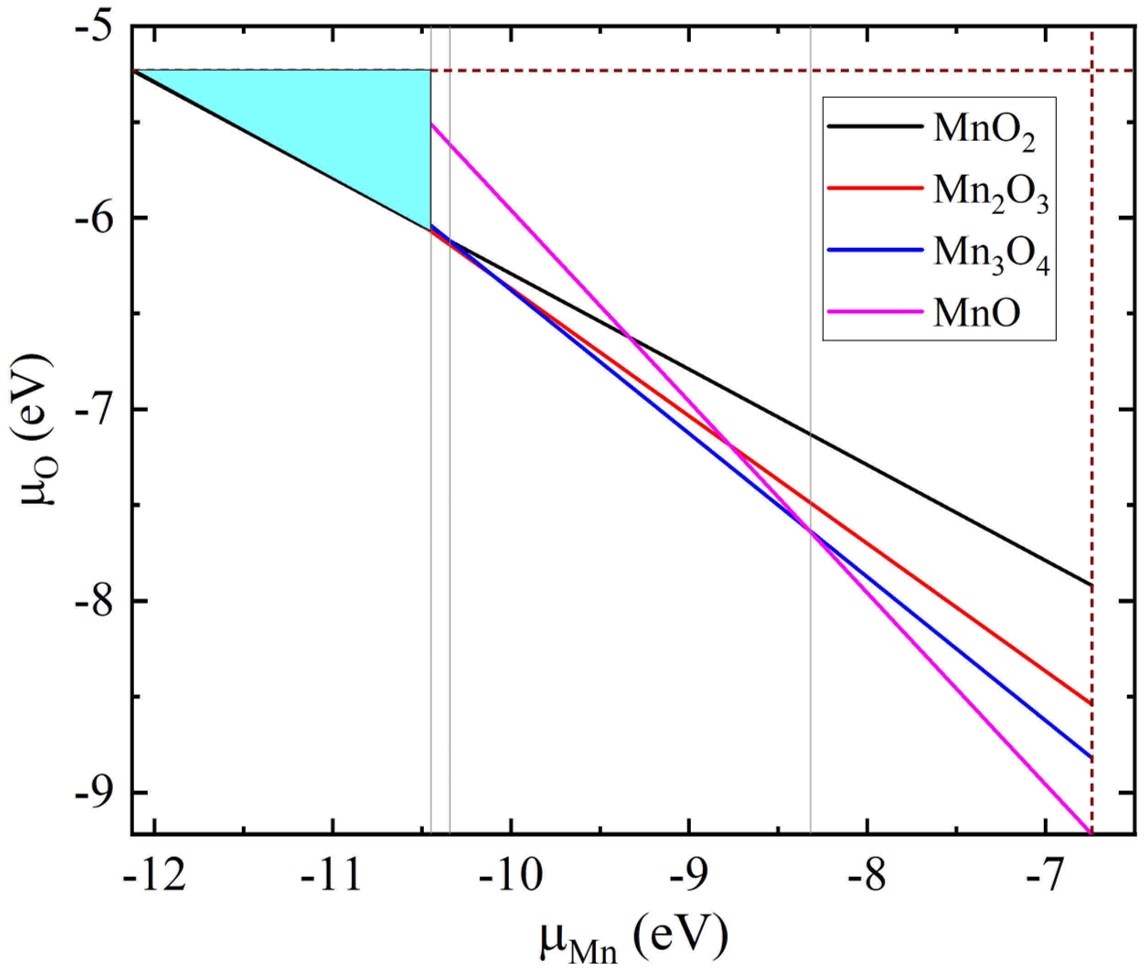}
    \caption{Chemical potential phase diagram (stability triangle) for $\alpha$-MnO$_2$ showing the allowed range of atomic chemical potentials $\mu_{\mathrm{Mn}}$ and $\mu_{\mathrm{O}}$. The dashed vertical and horizontal lines represent the upper limits set by the elemental phases of Mn and O, respectively. Thermodynamic stability of $\alpha$-MnO$_2$ is achieved when $\mu_{\mathrm{Mn}}$ and $\mu_{\mathrm{O}}$ lie within the cyan triangular region (black boundary). The boundary lines corresponding to competing phases $\alpha$-Mn$_2$O$_3$, Mn$_3$O$_4$, and MnO are shown in red, blue, and magenta, respectively.}
    \label{fig6}
\end{figure}

O vacancy (V\(_\text{O}\)) is modeled at two non-equivalent lattice O sites, with the more stable configuration favored by 364~meV [ref. Figs.~S4(d), S4(e)]. Removal of an O atom leaves excess electrons, which localize on two adjacent Mn sites, reducing them from Mn(IV) to Mn(III), as depicted in Fig.~\ref{fig6}(e), giving rise to a net magnetic moment of 2~$\mu_B$. A third Mn ion relaxes toward a neighboring O atom, forming a short Mn–O bond (1.82~\AA). The PDOS and the band structure reveal doubly occupied Mn-\textit{e}\(_g\) defect states inside the band gap, along with an unoccupied hybrid Mn-\textit{e}\(_g\)–O-2\textit{p} state [cf. Figs. \ref{fig4}(d) and S3(f)]. Notably, one of the occupied defect states lies very close to the VBM, suggesting it is unlikely to contribute to thermally activated ionization [cf. Fig. S3(f)]. The corresponding charge densities [Figs.~\ref{fig5}(e)-\ref{fig5}(f)] indicate strong electron localization on the reduced Mn sites and hole density on the third Mn ion, giving rise to both donor and acceptor states.

\subsubsection{Antisites}

In the Mn\(_\text{O}\) antisite, substitution of Mn at an O site causes the defect to relax toward the tunnel center, forming three Mn–O bonds of approximately 2.1~\AA{} [cf. Fig.~S4(f)]. The resulting excess electrons localize on four neighboring lattice Mn ions near V\(_\text{O}\), reducing them from Mn(IV) to Mn(III), while the substituted Mn remains unaltered in its electronic configuration. The PDOS and the band structure reveal spin-split occupied defect states within the gap, comprising three spin-majority (two Mn-\textit{e}\(_g\) and one Mn-\textit{t}\(_{2g}\)) orbitals and one spin-minority Mn-\textit{e}\(_g\) orbital hybridized with O-2\textit{p} orbitals [cf. Figs.~\ref{fig4}(d) and Fig. S3(d)]. Several unoccupied defect states are also observed near the CBM, although they are unlikely to contribute significantly to conduction. For clearer visualization, the corresponding band structure is shown in Fig. S3(d). The spin-polarized charge distributions of two occupied defect states near the Fermi level are shown in Figs.~\ref{fig5}(g) and \ref{fig5}(h). These reveal strong electron localization on the three reduced lattice Mn sites adjacent to the O-vacancy, as well as on a fourth reduced lattice Mn site in a neighboring octahedron, resulting in a net magnetic moment of 7 $\mu_B$.

For the O\(_\text{Mn}\) antisite, where an O substitutes a Mn site forms bond with two neighboring lattice O atoms of 1.41~\AA{} bond length each, resembling a peroxide-like O\(_2^{2-}\) species [cf. Fig.~S4(g)]. The PDOS and the band structure exhibit spin-split antibonding states near the Fermi level, originating from diatomic pp* hybridization [cf. Figs.~\ref{fig4}(g) and Fig. S3(g)]. As shown in Fig. S3(g), two occupied and two unoccupied states emerge close to the Fermi level, significantly reducing the band gap to approximately 240~meV and indicating the presence of shallow defect states. The spin densities of two distinct unoccupied states, illustrated in Figs.~\ref{fig5}(i) and \ref{fig5}(j), show hole delocalization across the defect O atom and surrounding lattice O atoms. This configuration results in a net magnetic moment of 3 $\mu_B$. The O\(_\text{Mn}\) antisite thus manifests hybrid features of both O\(_\text{i}\) and V\(_\text{Mn}\), but with comparatively less delocalized electronic states.

\begin{table}[ht]
\caption{\label{table2}
Calculated formation enthalpy $\Delta H_f$ (in eV) for various Mn-O phases using PBE+$U$ and HSE06 functionals, compared to the experimental measurements taken from Ref. \cite{fritsch1998thermochemistry} and theoretical values from Ref. \cite{franchini2007ground}. We cannot find reliable experimental data for $\Delta H_f$ for $\alpha$-MnO$_2$.}
\begin{ruledtabular}
\begin{tabular}{lcccccc}
 &  Present work & & & & Literature &\\
 &  PBE+$U$ & HSE06 & & PBE+$U$ & HSE06 & Expt.\\
\hline
MnO & -3.96 & -3.90 & &-3.92 & -3.83 & -3.99\\
$\alpha$-Mn$_2$O$_3$ & -9.74 & -9.54 & & -9.39 & -9.03 & -9.93\\
Mn$_3$O$_4$  & -13.9 & -14.0 & & -13.7 & -13.8 & -14.4\\
$\beta$-MnO$_2$ & -4.82 & -4.72 & & -4.51 & -4.47 & -5.39\\
$\alpha$-MnO$_2$ & -4.95 & -4.75 & & - & - & - \\
\end{tabular}
\end{ruledtabular}
\end{table}

\subsection{Phase stability}

For the thermodynamic stability of $\alpha$-MnO$_2$, the chemical potentials must satisfy the following condition:

\begin{equation}
    \mu_{\mathrm{Mn}} + 2\mu_{\mathrm{O}} = \Delta H_f(\mathrm{MnO_2}),
\label{eq2}
\end{equation}

where $\Delta H_f(\mathrm{MnO_2})$ is the formation enthalpy of $\alpha$-MnO$_2$, calculated to be $-4.95$ ($-4.75$) eV using the PBE+$U$ (HSE06) method. The reference chemical potentials $\mu_{\mathrm{Mn}}$ and $\mu_{\mathrm{O}}$ correspond to bulk Mn in its $\delta$ phase and the energy per atom of an O$_2$ molecule in its spin-polarized triplet ground state, respectively. These reference limits are indicated by the vertical and horizontal dashed lines in Fig.~\ref{fig6}. To prevent the formation of competing manganese oxide phases, such as $\alpha$-Mn$_2$O$_3$, Mn$_3$O$_4$, and MnO, the chemical potentials must also satisfy

\begin{equation}
    x\mu_{\mathrm{Mn}} + y\mu_{\mathrm{O}} \leq \Delta H_f(\mathrm{Mn}_x\mathrm{O}_y),
\label{eq3}
\end{equation}

where $\Delta H_f(\mathrm{Mn}_x\mathrm{O}_y)$ are the experimental formation enthalpies of the competing Mn$_x$O$_y$ phases listed in Table \ref{table2}. The formation energies obtained using PBE+$U$ and HSE06 are consistent with previous theoretical results~\cite{franchini2007ground} and are in good agreement with the experimental data reported in Ref.~\cite{fritsch1998thermochemistry}. Since the experimental formation enthalpy of $\alpha$-MnO$_2$ is not available, the experimental value of $\beta$-MnO$_2$ is used to estimate the lower bounds of $\mu_{\mathrm{Mn}}$ and $\mu_{\mathrm{O}}$, shown by the black solid line in Fig.~\ref{fig6}. The calculated $\Delta H_f(\mathrm{MnO_2})$ for $\beta$-MnO$_2$ is $-4.82$ ($-4.72$) eV using PBE+$U$ (HSE06), compared to the experimental value of $-5.39$ eV~\cite{fritsch1998thermochemistry}. Both functionals yield similar formation energies, consistent with earlier reports~\cite{kitchaev2016energetics}. Using the constraints given by Eqs.~(\ref{eq2}) and (\ref{eq3}), the upper bound of $\mu_{\mathrm{Mn}}$ and the corresponding lower bound of $\mu_{\mathrm{O}}$ are determined. Beyond these limits, the formation of competing phases such as $\alpha$-Mn$_2$O$_3$ becomes favorable. The resulting thermodynamic stability region of $\alpha$-MnO$_2$ is shown as the cyan area in the phase diagram in Fig.~\ref{fig6}.

\subsection{\label{sec:level2}Finite-size effects in charged defects}

\subsubsection{FNV and extended-FNV correction schemes}
Finite-size effects arising from spurious electrostatic interactions between periodically repeated charged defects are corrected using the electrostatics-based Freysoldt-Neugebauer-Van de Walle (FNV) scheme~\cite{freysoldt2009fully}. In this approach, a Gaussian model charge together with a static dielectric constant is employed to evaluate the electrostatic energy $E_{\mathrm{model}}(\Omega)$ for supercells of varying volume $\Omega$, which is subsequently extrapolated to the dilute limit to obtain the isolated defect energy $E_{\mathrm{iso}}$.

\begin{figure}[ht]
\centering
\includegraphics[width=1\linewidth]{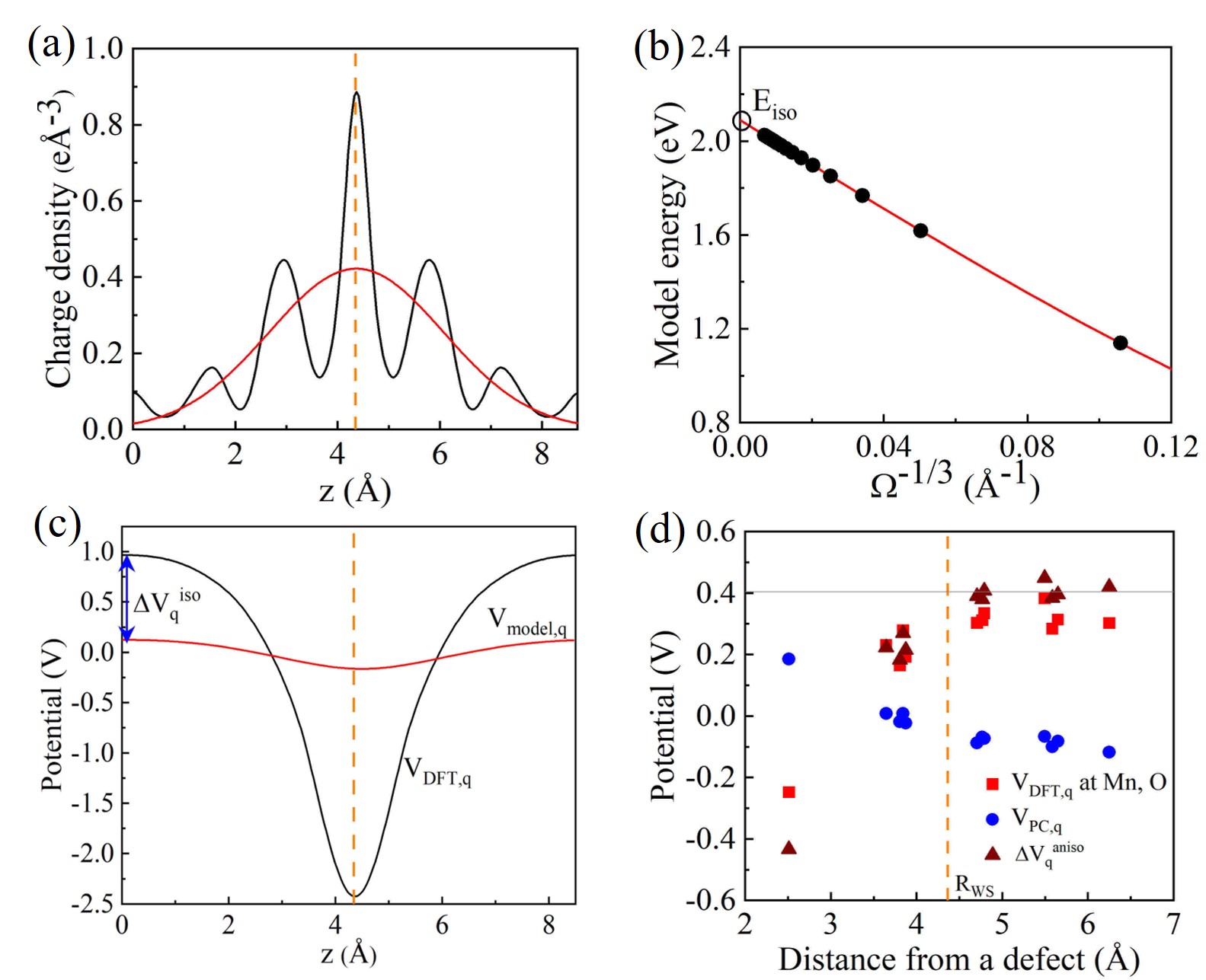}
\caption{(a) Plane-averaged defect charge densities computed using DFT (black solid line) and the model charge (red solid line) along the $c$-axis. (b) Electrostatic energies (filled circles) and the corresponding third-order polynomial fit (red solid line) for the (2$^+$) charged defect as a function of the inverse cube root of the supercell volume ($\Omega$). The extrapolated value (E$_{\text{iso}}$) at $\Omega^{-1/3} = 0$ is indicated by an open circle. (c) Plane-averaged model potential, $V_\text{model,q}$ (red) and DFT difference potential (black). (d) Electrostatic potentials $V_\text{DFT,q}$, $V_{\text{PC,q}}$, and $\Delta$$V_{\text{q}}^\text{aniso}$ evaluated at the atomic positions in the 1$\times$1$\times$3 $\alpha$-MnO$_2$ supercell with a Mn\(_\text{i}\) defect. The spatial region used for averaging $\Delta$$V_{\text{q}}^\text{aniso}$ is outside of the sphere with radius $R_\text{WS}$. The defect center is shown by the orange dashed line.}
\label{fig7}
\end{figure}

The model electrostatic energy for the doubly positively charged Mn interstitial (Mn$_{\mathrm{i}}^{2+}$) is evaluated using a Gaussian charge distribution with a width of 1.7~\AA, chosen to reproduce the spatial extent of the DFT-computed defect charge density [cf. Fig.~\ref{fig7}(a)]. Starting from a $1\times1\times3$ supercell, the cell is uniformly scaled, and $E_{\mathrm{model}}$ is computed as a function of $\Omega$ [cf. Fig.~\ref{fig7}(b)]. The resulting data are fitted to a third-order polynomial and extrapolated to the infinite-size limit, yielding $E_{\mathrm{iso}} = 2.09$~eV. The difference between $E_{\mathrm{iso}}$ and $E_{\mathrm{model}}$ for the $1\times1\times3$ supercell defines the lattice correction $E_{\mathrm{lat}}$, which accounts for long-range Coulomb interactions.

In addition, electrostatic corrections are evaluated using the extended-FNV (eFNV) scheme following Kumagai and Oba~\cite{kumagai2014electrostatics}, which is particularly suitable for systems with anisotropic dielectric screening. In contrast to the original FNV approach, the eFNV method employs a point-charge (PC) model, enabling a rigorous treatment of long-range Coulomb interactions within an anisotropic dielectric medium. Within this formalism, the correction to the defect formation energy in Eq. \ref{eq1} is expressed as
\begin{equation}
    E_\mathrm{corr}^{\mathrm{FNV/eFNV}} = E_{\mathrm{FNV}} - q\,\Delta V_{\text{q}}^{\mathrm{iso/aniso}}.
\end{equation}

Here, $E_{\mathrm{FNV}}$ corresponds to the lattice correction $E_{\mathrm{lat}}$ in the FNV scheme and to the point-charge correction energy $E_{\mathrm{PC}}$ in the eFNV scheme. The term $\Delta V_{q}^{\mathrm{iso/aniso}}$ represents the potential alignment-like correction. In the FNV scheme, this term is obtained by comparing the planar-averaged model potential ($V_{\mathrm{model},q}$) with the DFT defect-induced potential ($V_{\mathrm{DFT},q}$), defined as the difference between the electrostatic potentials of the defective and pristine supercells, $V_{\mathrm{DFT},q} = V_q - V_{\mathrm{bulk}}$ as shown in Fig.~\ref{fig7}(c). In the eFNV scheme, $\Delta V_{q}^{\mathrm{aniso}}$ is evaluated using atomic site electrostatic potentials by comparing the PC potential $V_{\mathrm{PC},q}$ with $V_{\mathrm{DFT},q}$. The PC potential at an arbitrary position $\mathbf{r}$ is computed using the generalized Ewald summation:
\begin{equation}
\begin{aligned}
V_\mathrm{PC,q}(\mathbf{r}) =
 \sum_{\mathbf{R}_i}
\frac{q}{\sqrt{|\boldsymbol{\varepsilon}|}}
\frac{\operatorname{erfc}\!\left(
\gamma \sqrt{(\mathbf{R}_i-\mathbf{r}) \cdot
\boldsymbol{\varepsilon}^{-1} \cdot (\mathbf{R}_i-\mathbf{r})}
\right)}
{\sqrt{(\mathbf{R}_i-\mathbf{r}) \cdot 
\boldsymbol{\varepsilon}^{-1} \cdot (\mathbf{R}_i-\mathbf{r})}} \\
 - \frac{\pi q}{\Omega \gamma^2}
+ \sum_{\mathbf{G}_i \ne 0}
\frac{4\pi q}{\Omega}
\frac{\exp\!\left(
-\mathbf{G}_i \cdot \boldsymbol{\varepsilon} \cdot \mathbf{G}_i
/ 4\gamma^2
\right)}
{\mathbf{G}_i \cdot \boldsymbol{\varepsilon} \cdot \mathbf{G}_i}
\, \cdot e^{i \mathbf{G}_i \cdot \mathbf{r}} ,
\end{aligned}
\end{equation}

where $\mathbf{R}_\text{i}$ and $\mathbf{G}_\text{i}$ denote lattice vectors in real and reciprocal space, respectively. The alignment-like term $\Delta V_{q}^{\mathrm{aniso}}$ is obtained by averaging $\Delta V_{\mathrm{PC},q}$ over atomic sites within a sampling region located outside a sphere inscribed within the Wigner-Seitz cell of radius $R_{\mathrm{WS}}$, thereby minimizing the influence of local structural relaxations. This procedure replaces the planar-averaging approach of the original FNV scheme and provides a more robust evaluation for systems exhibiting significant atomic displacements. As illustrated in Fig.~\ref{fig7}(d), the atomic site potentials $V_{\mathrm{DFT},q}$, $V_{\mathrm{PC},q}$, and $\Delta V_{q}^{\mathrm{aniso}}$ exhibit significant spatial variation in the vicinity of the defect, but converge to a well-defined constant value in the region beyond $R_{\mathrm{WS}}$, enabling an unambiguous determination of the alignment correction.

The total correction thus comprises a long-range Coulomb contribution ($E_{\mathrm{FNV}}$) and a short-range potential alignment-like term ($-q\,\Delta V_{q}^{\mathrm{iso/aniso}}$), which together remove spurious electrostatic interactions under periodic boundary conditions. The resulting correction energies for all defects considered in this work, obtained using both FNV and eFNV schemes, are summarized in Table~S1.

\subsubsection{Finite-size convergence and role of dielectric anisotropy in charge corrections}

\begin{figure}[ht]
    \centering
    \includegraphics[width=1\linewidth]{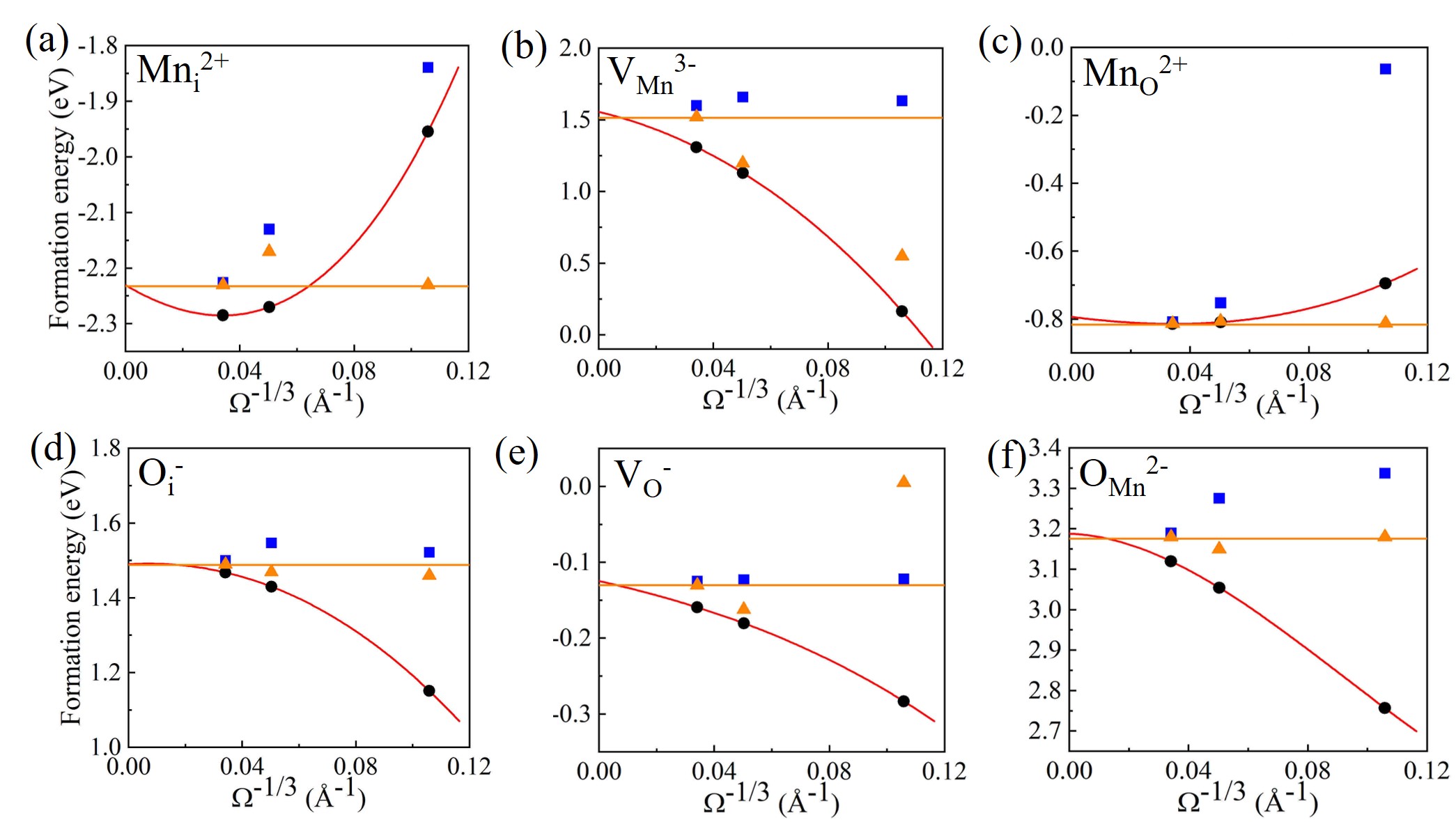}
    \caption{Calculated formation energies of (a) Mn$_\text{i}^{2+}$, (b) V$_\text{Mn}^{3-}$, (c) Mn$_\text{O}^{2+}$, (d) O$_\text{i}^{-}$, (e) V$_\text{O}^{-}$, and O$_\text{Mn}^{2-}$ in $\alpha$-MnO$_2$ as a function of the inverse cube root of the  supercell volume ($\Omega$). The horizontal axis represents the formation energy calculated with the largest supercell and extended FNV corrections. The uncorrected formation energies (black solid circles) are fitted with a third order polynomial function (red line). The isotropic FNV corrections and anisotropic FNV corrections are represented by blue squares and orange triangles.}
    \label{fig8}
\end{figure}

To further evaluate finite-size corrections, we compute the formation energies of selected charged defects (Mn$_i^{2+}$, V$_{\mathrm{Mn}}^{3-}$, Mn$_\mathrm{O}^{2+}$, O$_i^{1-}$, V$_\mathrm{O}^{1-}$, and O$_\mathrm{Mn}^{2-}$) using $1 \times 1 \times 3$ (72-atom), $2 \times 2 \times 7$ (672-atom) and $3 \times 3 \times 10$ (2160-atom) supercells as shown in Fig. \ref{fig8}. These supercells are constructed to ensure near-isotropic expansion, with the $3 \times 3 \times 10$ supercell approaching a cubic geometry. The uncorrected energies are extrapolated to the dilute limit using $a\Omega^{-1} + b\Omega^{-1/3} + c$, where $\Omega$ is the supercell volume. Without corrections, strong size dependence is observed, whereas finite-size correction significantly improves convergence.

The role of dielectric anisotropy is examined through a systematic comparison between the isotropic FNV and anisotropic eFNV correction schemes. The relative performance of the two approaches is found to depend sensitively on the specific defect and charge state. In particular, the eFNV scheme yields improved convergence with increasing supercell size for defects such as Mn$_\text{i}^{2+}$ and Mn$_\mathrm{O}^{2+}$, primarily due to differences in the long-range electrostatic correction term, while the contribution from potential alignment remains comparatively small [cf. Table~S1]. The strong sensitivity of Mn$_\text{i}$ and Mn$_\mathrm{O}$ to the dielectric treatment reflects their anisotropic charge distribution along the tunnel structure, thereby necessitating a tensorial dielectric description. In contrast, V$_{\mathrm{Mn}}^{3-}$ and V$_{\mathrm{O}}^{1-}$ exhibit comparatively better convergence behavior with the isotropic FNV scheme, whereas O$_\text{i}^{1-}$ and O$_\mathrm{Mn}^{2-}$ show only weak sensitivity to the choice of correction method, with differences remaining within $\sim$0.1~eV. These results indicate that no single correction scheme universally provides the best convergence for all defects. Therefore, in the present work, the correction methodology for each defect configuration is selected based on the supercell convergence behavior, with the corrected formation energies referenced to the largest supercell calculations where the differences between the FNV and eFNV schemes become substantially reduced.

The charge-state dependence of the correction is analyzed in Fig.~S5. While an overall quadratic dependence on $q$ is observed, the data are better described by $E_{\text{corr}} = aq^2 + bq + c$, where the linear term arises from potential alignment. This contribution is significant for low charge states and leads to deviations from simple proportional scaling (e.g., between $q=+1$ and $q=+2$). For Mn$_i^{q}$, partial cancellation between quadratic and linear terms results in a relatively smaller correction for $q=+2$. This behavior is consistently observed in both FNV and eFNV schemes, indicating that it is intrinsic to the correction formalism.

Based on the above results, the eFNV scheme is employed for all defects except V$_{\mathrm{Mn}}$ and V$_\mathrm{O}$, for which the standard FNV correction is adequate.

\subsection{Defect formation energies}

\begin{figure}[ht]
    \centering
    \includegraphics[width=0.7\linewidth]{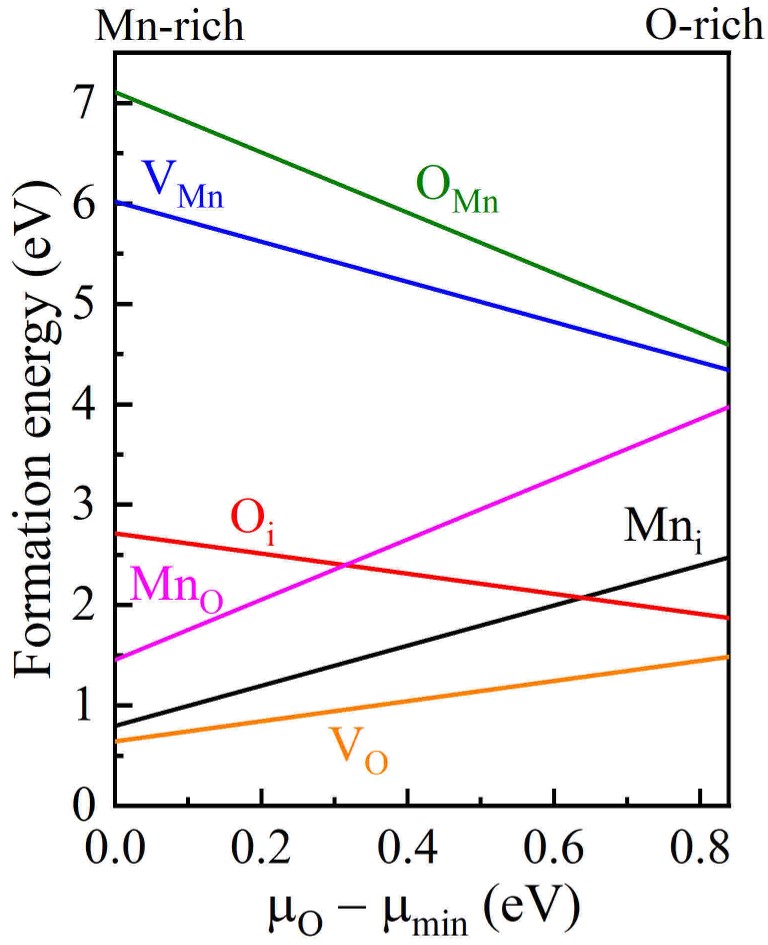}
    \caption{Calculated formation energies of native point defects in the neutral state as a function of the oxygen chemical potential (\(\mu_\text{O}\)), referenced to the minimum oxygen chemical potential (\(\mu_{\min}\)), in the \(1 \times 1 \times 3\) $\alpha$-MnO$_2$ supercell. The range of \(\mu_\text{O}\) corresponds to the thermodynamic stability limits of the host compound.}
    \label{fig9}
\end{figure}

The formation energies of native point defects in $\alpha$-MnO$_2$ are evaluated as a function of oxygen chemical potential under Mn-rich and O-rich conditions as shown in Fig.~\ref{fig9}. To assess convergence, calculations are performed using both $1 \times 1 \times 3$ (72-atom) and $2 \times 2 \times 7$ (672-atom) supercells (Table~\ref{table3}). The formation energies differ by only a fraction of an eV, indicating that the smaller supercell is sufficient for describing native defects. The corresponding magnetic moments are also listed in Table~\ref{table3}.

\begin{table}[ht]
\caption{Defect formation energies (in eV) for native point defects in $\alpha$-MnO$_2$ computed using DFT+$U$ in two different supercell sizes: $1 \times 1 \times 3$ (72 atoms) and $2 \times 2 \times 7$ (672 atoms), under Mn-rich and O-rich conditions. The last column lists the net magnetic moment ($\mu$) for each defect.}
\label{table3}
\setlength{\tabcolsep}{6pt}
\renewcommand{\arraystretch}{1.2}
\begin{tabular}{lccccc}
\hline\hline
\multirow{2}{*}{\shortstack{Defect\\type}} & \multicolumn{2}{c}{$1 \times 1 \times 3$} & \multicolumn{2}{c}{$2 \times 2 \times 7$} & \multirow{2}{*}{\shortstack{$\mu$\\($\mu_B$)}}\\
\cline{2-5}
& Mn-rich & O-rich & Mn-rich & O-rich & \\
\hline
Mn$_\text{i}$   & $0.798$ & $2.48$ & $0.955$ & $2.63$ & 7 \\
O$_\text{i}$    & $2.71$  & $1.87$ & $2.78$  & $1.94$ & 0 \\
V$_\text{Mn}$   & $6.02$ & $4.34$ & $6.07$ & $4.40$ & 5 \\
V$_\text{O}$    & $0.643$ & $1.48$ & $0.679$ & $1.52$ & 2 \\
Mn$_\text{O}$   & $1.46$ & $3.97$ & $1.59$ & $4.10$ & 7 \\
O$_\text{Mn}$   & $7.11$ & $4.59$ & $7.11$ & $4.59$ & 3 \\
\hline\hline
\end{tabular}
\end{table}

V$_\text{O}$ is the most stable neutral defect, followed by Mn$_\text{i}$ and O$_\text{i}$ under Mn-rich conditions and O-rich conditions, respectively. The uncorrected formation energies for various charge states are shown in Figs.~S6 and S7, while the FNV and eFNV corrected results are presented in Figs.~S8-S11, with numerical values summarized in Table~S2. The corresponding charge transition levels (CTLs) are listed in Table~\ref{table4}. The differences in CTLs obtained using the FNV and eFNV schemes arise from the variation in their respective correction energies. Notably, relatively large deviations are estimated for most defects, whereas the O$_\text{i}$ and O$_\mathrm{Mn}$ defects exhibit only small differences within $\sim$0.1~eV. Additionally, finite-size effects for charged defects are assessed using the larger $2 \times 2 \times 7$ supercell. The corrected formation energies differ by less than $\sim$0.1-0.15~eV compared to the smaller cell [cf. Table S2], and CTLs shift by less than $\sim$0.1~eV [cf. Table S3]. The qualitative features, including the amphoteric nature of V$_{\mathrm{O}}$, remain unchanged, confirming the reliability of the $1 \times 1 \times 3$ supercell.

\begin{table}[ht]
\centering
\caption{Charge transition levels (in eV) for native point defects in \(1 \times 1 \times 3\) $\alpha$-MnO$_2$ supercell corrected using FNV (extended FNV) approach, referenced to the VBM.}
\label{table4}
\setlength{\tabcolsep}{10pt}
\renewcommand{\arraystretch}{1.2}
\setlength{\tabcolsep}{3pt}  
\begin{tabular}{lccccc}
\hline\hline
\multirow{2}{*}{\shortstack{Defect\\type}} & \multicolumn{5}{c}{Charge transition levels} \\
\cline{2-6}
& (2+/1+) & (1+/0) & (0/1$-$) & (1$-$/2$-$) & (2$-$/3$-$) \\
\hline
Mn$_\text{i}$   & 1.24 & 1.39 &       &       &       \\
&  (1.47) &   &       &       &       \\
\hline
O$_\text{i}$    &       & 0.328 & 1.17 &       &       \\
    &       & (0.358) & (1.11) &       &       \\
\hline
V$_\text{Mn}$   &       &       &       & 0.636 & 1.31 \\
 &       &       &       & (0.277) & (0.699) \\
\hline
V$_\text{O}$    &  & 0.228 & 0.755 &       &       \\
    &  & (0.208) & (0.885) &       &       \\
\hline
Mn$_\text{O}$   & 0.263 & 1.25 &       &       &       \\
   & (0.676) &  &       &       &       \\
\hline
O$_\text{Mn}$   &       &       & 0.512 & 1.28 &     \\
   &       &       & (0.407) & (1.23) &     \\
\hline\hline
\end{tabular}
\end{table}

\subsubsection{Interstitials}

The Mn$_{\mathrm{i}}$ defect exhibits a relatively low formation energy of 0.798~eV in the neutral charge state under Mn-rich conditions [cf. Fig.~\ref{fig9}]. Upon inclusion of FNV corrections, the negative-$U$ behavior associated with the (1$^+$) charge state in the uncorrected formation energies is removed [cf. Figs. S8(a) and S9(a)]. As a result, the defect stabilizes in distinct (2$^+$), (1$^+$), and (0) charge states across the band gap. However, within the eFNV correction scheme, the defect exhibits negative-$U$ behavior involving the neutral charge state, which favors direct transitions between the ionized leading to ionized (2$^+$) and (1$^+$) configurations. Nevertheless, the (1$^+$) charge state remains thermodynamically stable. The thermodynamic transition level $\epsilon(2^+/1^+)$ is located 44~meV below the CBM, indicating that Mn$_{\mathrm{i}}$ behaves as a shallow double donor. Similar behavior has been reported for interstitial defects in TiO$_2$ and SnO$_2$~\cite{kilicc2002origins,na2006first}. Under Mn-rich conditions, the corrected formation energies of the (1$^+$) and (2$^+$) charge states are $-0.755$~eV and $-2.23$~eV, respectively [cf. Fig. S10(a)]. Under O-rich conditions, Mn$_{\mathrm{i}}$ continues to exhibit comparatively low formation energy in the (2$^+$) charge state [cf. Fig. S11(a)], while remaining thermodynamically accessible in the (1$^+$) state.

The O$_\text{i}$ defect allows for (1$^+$), (0), and (1$^-$) charge states within the band gap [cf. Figs. S8(b) and S9(b)]. Under O-rich conditions, the neutral O$_\text{i}$ has a low formation energy of 1.87~eV [cf. Fig.~\ref{fig7}]. Using model charge calculation, the correction to formation energy is 0.245 eV, while PC correction energy is 0.131 eV . After accounting for potential alignment, E$_\text{corr}^\text{FNV}$ is calculated to be 0.157 eV and 0.371 eV for the (1$^+$) and (1$^-$) states, respectively, comparable to the corresponding E$_\text{corr}^\text{eFNV}$ values of 0.126 eV and 0.312 eV [cf. Table S1]. The charge transition levels $\epsilon(1^+/0)$ and $\epsilon(0/1^-)$ are located at 1.16~eV below the CBM and 1.11~eV above the VBM, respectively, indicating amphoteric behavior.

\subsubsection{Vacancies}

The Mn vacancy (V$_\text{Mn}$) has high formation energies of 4.34~eV and 6.02~eV under O-rich and Mn-rich conditions, respectively, due to the need to break six Mn-O bonds [cf. Fig.~\ref{fig9}]. The stable charge states are (1$^-$), (2$^-$), and (3$^-$) as indicated in Figs. S8(c) and S9(c).  The defect exhibits negative-$U$ behavior associated with the neutral state, which is destabilized compared to its ionized forms. As a result, V$_\text{Mn}$ is predicted to always be ionized. The FNV-corrected transition levels, $\epsilon(1^-/2^-)$ and $\epsilon(2^-/3^-)$, are located 0.636~eV and 1.31~eV above the VBM, respectively. Therefore, V$_\text{Mn}$ can behave not only as a shallow acceptor but also as an electron trap near the CBM. FNV corrections to the formation energy are 0.037 eV, 0.523 eV, and 1.47 eV for the (1$^-$), (2$^-$), and (3$^-$) states, respectively [cf. Table S1].

The O vacancy (V$_\text{O}$) has the lowest formation energy under both Mn-rich and O-rich conditions in the neutral state, with formation energies of 0.643~eV and 1.48 eV, respectively, as shown in Fig.~\ref{fig9}]. Stable charge states include (1$^+$), (0), and (1$^-$) within the band gap region [cf. Figs. S8(d), S9(d)]. The transition level $\epsilon(1^+/0)$ lies 1.29~eV below the CBM, respectively, while $\epsilon(0/1^-)$ is positioned 0.755~eV above the VBM, which act as compensating centers. The presence of both donor-like and acceptor-like charge states confirms the amphoteric nature of V$_\text{O}$. Most first-principles studies of oxygen vacancies in transition-metal oxides have primarily considered only positively charged and neutral states (V$^{2+}$, V$^{+}$, and V$^{0}$). In TiO$_2$, V$_\mathrm{O}$ acts as a double donor and remains ionized across the Fermi level, with no amphoteric behavior reported~\cite{na2006first}. In SnO$_2$, V$_\mathrm{O}^{2+}$ and Sn$_\mathrm{i}^{4+}$ are reported as dominant donor-type defects exhibiting shallow states~\cite{kilicc2002origins}. In both cases, finite-size corrections were not explicitly included. In contrast, Ramo \textit{et al.}~\cite{munoz2007spectroscopic} considered multiple charge states, including negative ones, in monoclinic HfO$_2$, suggesting that V$_\mathrm{O}$ can also act as an electron trap. However, the position of these defect levels relative to the conduction-band edge may be underestimated, possibly due to the absence of finite-size corrections. Overall, the amphoteric behavior of V$_\mathrm{O}$ remains less explored in most oxides and sensitive to computational treatment.

To assess the sensitivity of defect levels to the Hubbard parameter ($U$), we evaluated the charge transition levels (CTLs) of V$_\text{O}$ for $U$ values in the range of 2-5 eV as shown in Fig. S12. The evolution of CTLs with $U$ reflects the increased localization of Mn-3$d$ states, which modifies their relative alignment with respect to the band edges. The donor transition level shifts progressively toward the VBM with increasing $U$, while the acceptor level moves toward the midgap region up to $U = 3.8$ eV and shifts back toward the CBM for larger $U$ values. Importantly, within the physically justified range ($U \le 3.8$ eV), both donor- and acceptor-type levels remain inside the band gap, indicating that the amphoteric character of V$_\text{O}$ is robust with respect to the choice of $U$.

\begin{figure*}[ht]
\centering
\includegraphics[width=0.6\textwidth]{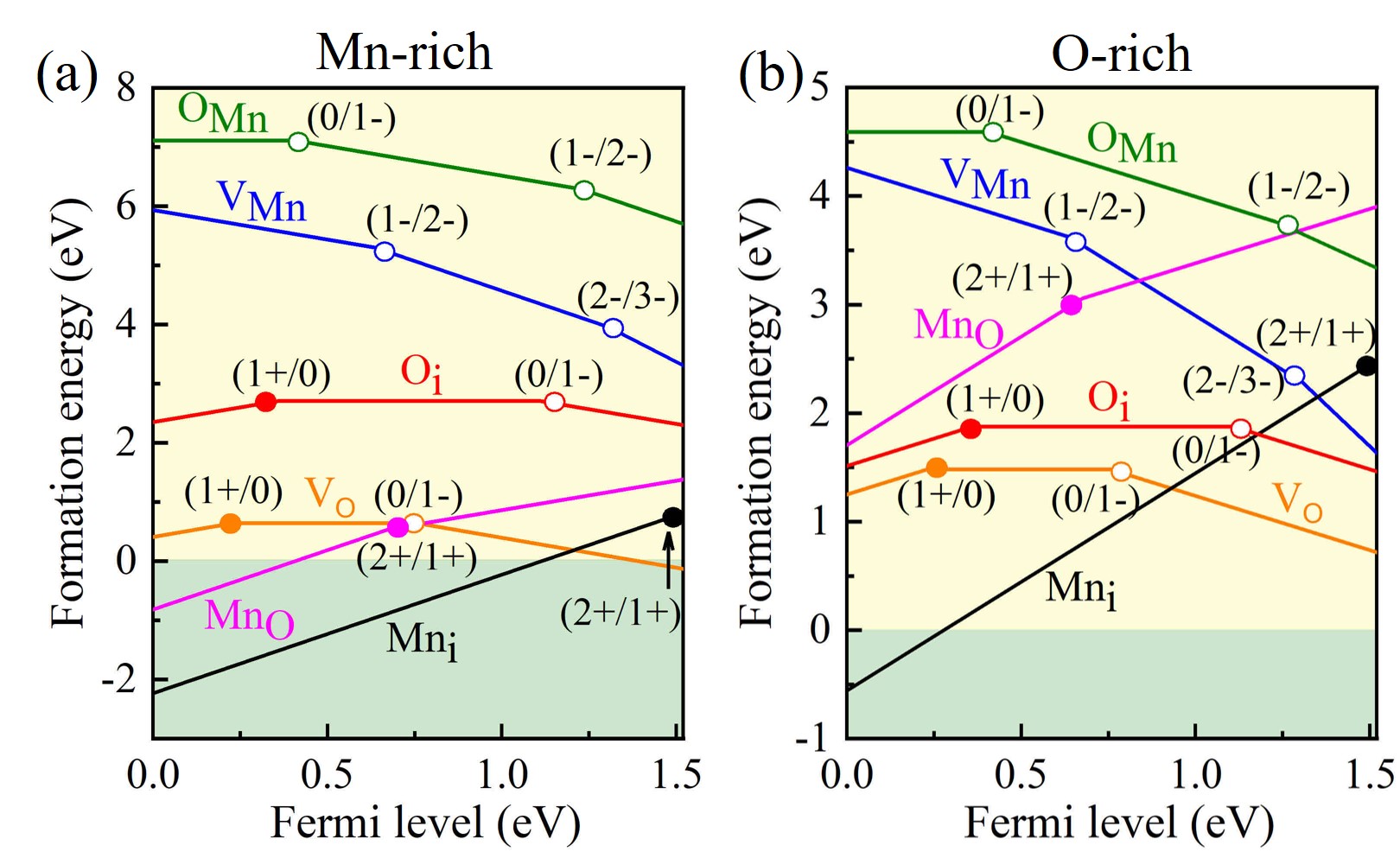}
\caption{(a), (b) Variation of formation energies for stable charge states of native point defects in \(1 \times 1 \times 3\) $\alpha$-MnO$_2$ supercell as a function of the Fermi level under Mn-rich and O-rich conditions. Charge transition levels are indicated in parentheses. The green (yellow) shaded region represents the regime of spontaneous (non-spontaneous) defect formation.}
\label{fig10}
\end{figure*}

\subsubsection{Antisites}

The Mn$_{\mathrm{O}}$ antisite exhibits a relatively low formation energy of 1.46~eV in the neutral charge state under Mn-rich conditions [cf. Fig.~\ref{fig9}]. Within the band gap, the defect stabilizes in the (0), (1$^+$), and (2$^+$) charge states upon inclusion of FNV corrections [cf. Figs. S8(e) and S9(e)]. Upon incorporating eFNV corrections, the defect exhibits negative-$U$ behavior associated with the neutral state. Consequently, the neutral state remains thermodynamically unstable, and the defect preferentially exists in ionized configurations, indicative of shallow donor character, similar to the Mn$_{\mathrm{i}}$ defect [cf. Figs. S10(e) and S11(e)]. The thermodynamic transition level, $\epsilon(2^+/1^+)$, is located 0.844~eV below the conduction band minimum (CBM), suggesting that the defect can act both as a compensating center and as a donor.

The O$_\text{Mn}$ antisite is found to be stable in the (0), (1$^-$), and (2$^-$) charge states within the gap [Figs. S8(f), S9(f)]. It has high formation energies of 4.59~eV under O-rich and 7.11~eV under Mn-rich conditions, owing to the disruption of multiple Mn-O bonds. The transition levels $\epsilon(0/1^-)$ and $\epsilon(1^-/2^-)$ (including eFNV corrections) are located 0.407~eV and 1.23~eV above the VBM, respectively, indicative of deep acceptor states. The formation energy corrections (E$_\text{corr}^\text{eFNV}$), including potential alignment, are -0.036~eV and 0.428~eV for the (1$^-$) and (2$^-$) states, respectively.

\subsection{Stability}

Figs.~\ref{fig10}(a) and \ref{fig10}(b) show the formation energies of isolated native defects in $\alpha$-MnO$_2$ as a function of the chemical potential (Fermi level) under Mn-rich and O-rich growth conditions, respectively. These plots identify the thermodynamically stable defect charge states and their corresponding charge transition levels over the accessible Fermi-level range. They also provide insight into the intrinsic defect thermodynamics that are relevant for extrinsic doping strategies. The defect formation energies are evaluated using the eFNV correction scheme for all defects except V$_{\mathrm O}$ and V$_{\mathrm{Mn}}$, for which the conventional FNV correction provides better convergence, as discussed in the previous sections. For completeness, formation-energy diagrams obtained using purely FNV and purely eFNV corrections are presented in Figs.~S13 and S14, respectively. Apart from small quantitative differences for Mn$_{\mathrm i}$ and Mn$_{\mathrm O}$, the relative stability of the defect charge states and the overall defect thermodynamics remain unchanged.

Energetically, Mn$_\mathrm{i}$ and V$_\mathrm{O}$ are the dominant native defects in $\alpha$-MnO$_2$. Mn$_\mathrm{i}$ remains ionized in the (2$^+$) and (1$^+$) charge states and introduces shallow donor levels, with the $\varepsilon(2^+/1^+)$ transition level located 44 meV below the CBM [cf. Fig. \ref{fig10}]. The negative formation energy of the (2$^+$) state indicates a strong tendency for spontaneous incorporation at tunnel sites in high concentrations, consistent with experimental observations~\cite{yuan2016influence,gao2008microstructures}. In contrast, V$_\mathrm{O}$ exhibits amphoteric behavior in $\alpha$-MnO$_2$, introducing both donor- and acceptor-like levels within the band gap. The (1$^-$) charge state of V$_\mathrm{O}$ becomes energetically favorable and shows negative formation energy under Mn-rich conditions. Unlike many binary oxides, where oxygen vacancies often act as shallow donors~\cite{kilicc2002origins,na2006first}, the oxygen vacancy in $\alpha$-MnO$_2$ forms comparatively deep donor levels.

The formation energy diagrams further reveal pronounced intrinsic defect compensation. Under both Mn-rich and O-rich conditions, Mn$_{\mathrm i}^{2+}$ possesses the lowest formation energy over a wide Fermi-level range near the VBM, whereas V$_{\mathrm O}^{-}$ becomes increasingly favorable as the Fermi level approaches the CBM. Consequently, any attempt to shift the Fermi level toward either band edge thermodynamically promotes the formation of compensating native defects. Such behavior indicates that native defects impose intrinsic constraints on the achievable Fermi-level position. To illustrate these compensation tendencies, we identify the Fermi-level positions at which the formation energies of Mn$_{\mathrm i}^{2+}$ and V$_{\mathrm O}^{-}$ become zero. These quantities should not be interpreted as equilibrium Fermi levels or rigorous doping limits and serve only as indicators of the onset of spontaneous donor or acceptor defect formation.

Under Mn-rich conditions, the formation energy of Mn$_{\mathrm i}^{2+}$ becomes zero $\sim$1.14 eV above the VBM, indicating that moving the Fermi level further toward the valence band edge would strongly favor spontaneous formation of donor-like Mn interstitials, thereby compensating acceptors. In contrast, the formation energy of V$_{\mathrm O}^{-}$ approaches zero $\sim$0.173 eV below the CBM, indicating that V$_\text{O}$ becomes thermodynamically accessible only when the Fermi level approaches the conduction band edge. These trends indicate that Mn$_\text{i}$ constitutes the dominant low-energy donor-type native defect under Mn-rich growth conditions, while V$_\text{O}$ acts as the corresponding compensating acceptor-like defect as the Fermi level approaches the conduction-band edge. Together, these defects illustrate the strong intrinsic self-compensation revealed by the calculated defect thermodynamics.

Under O-rich conditions, the formation energies of Mn-related defects (Mn$_{\mathrm i}$, Mn$_{\mathrm O}$ and V$_{\mathrm O}$) increase significantly, whereas those of O-related defects (V$_{\mathrm{Mn}}$, O$_{\mathrm{Mn}}$ and O$_{\mathrm i}$) decrease. Consequently, the onset of spontaneous Mn$_{\mathrm i}$ formation shifts closer to the VBM ($\sim$0.289 eV), reflecting a reduced tendency for donor compensation under O-rich conditions. At the same time, V$_{\mathrm O}^{-}$ remains energetically accessible near the conduction band edge. These results demonstrate that the balance between donor- and acceptor-type native defects is highly sensitive to the chemical environment. However, the present calculations alone do not establish the equilibrium carrier type or the feasibility of intrinsic n- or p-type conductivity, which would require a full charge-neutrality analysis. Rather, they identify the dominant native compensating defects that must be considered in future extrinsic doping strategies for $\alpha$-MnO$_2$.

Other defects such as V$_{\mathrm{Mn}}$ and O$_{\mathrm{Mn}}$ exhibit relatively higher formation energies, and thus their equilibrium concentrations remain low. Nevertheless, V$_{\mathrm{Mn}}$ remains ionized across the entire Fermi level range, with V$_{\mathrm{Mn}}^{-}$ acting as a shallow acceptor. Although these defects are not thermodynamically dominant, they may become relevant under non-equilibrium growth conditions or in the presence of suitable extrinsic dopants.

To elucidate the low formation energies of Mn$_\text{i}^{2+}$ and V$_\text{O}^{-}$, we analyze the associated structural relaxations in $\alpha$-MnO$_2$. Mn$_\text{i}^{2+}$ is readily accommodated within the tunnel structure with minimal lattice distortion. In contrast, a neutral Mn atom, due to its larger effective size, would introduce significant strain, resulting in a higher formation energy. Ionized Mn species are therefore sufficiently compact to fit within the tunnel, which explains the relative stability of charged Mn interstitials, with Mn$_\text{i}^{2+}$ being the most favorable configuration.

Removal of an O atom coordinated to three Mn ions introduces two excess electrons that localize on two Mn sites, leaving the third Mn relatively electron-deficient with an unoccupied defect state [cf. Fig. \ref{fig5}]. This uneven charge distribution induces asymmetric relaxation: the Mn associated with the hole state shifts outward significantly ($\sim$0.09 \AA), while the other two Mn show minor displacements toward the defect ($\sim$0.015 \AA). For V$_{\mathrm{O}}^{+}$, the relaxation becomes more symmetric, with the two Mn ions moving slightly away from the defect centre ($\sim$0.035 \AA) and the third Mn moving slightly toward the vacancy ($\sim$0.021 \AA). In contrast, V$_{\mathrm{O}}^{-}$ exhibits strongly asymmetric distortion: the two Mn ions remain nearly unchanged, while the Mn associated with the hole state moves markedly towards the defect site ($\sim$0.152 \AA). The pronounced inward displacement reflects stabilization of the localized hole state, driving charge redistribution and structural distortion that underpin the amphoteric behavior of V$_{\mathrm{O}}$.

\begin{figure}[ht]
\centering
\includegraphics[width=1\linewidth]{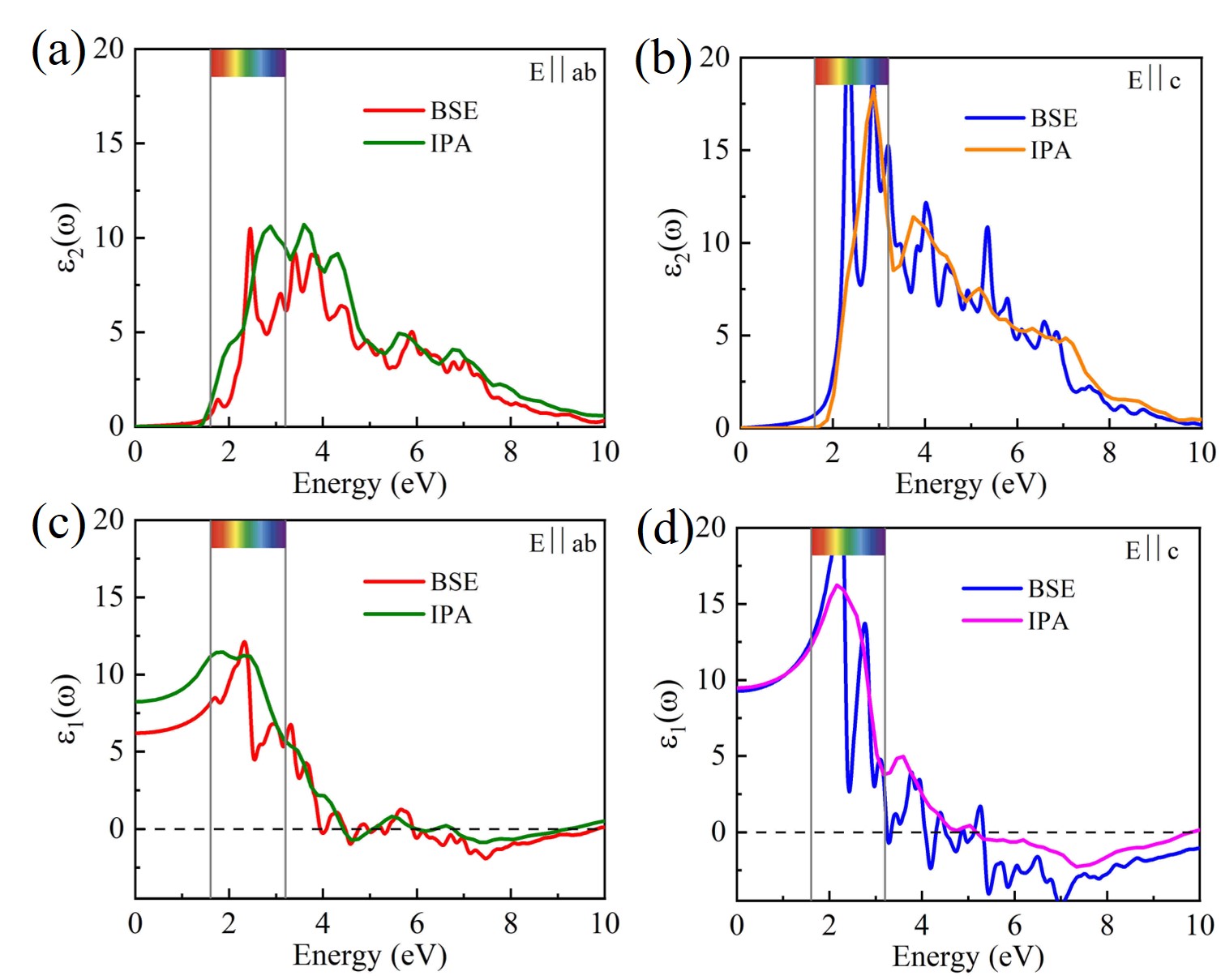}
\caption{(a),(b) Imaginary part of dielectric constant $\varepsilon_2(\omega)$ and (c),(d) real part of dielectric constant $\varepsilon_1(\omega)$ for \(1 \times 1 \times 3\) $\alpha$-MnO$_2$ supercell, calculated using IPA and GW+BSE level of theory.}
\label{fig11}
\end{figure}

\begin{figure*}[ht]
\centering
\includegraphics[width=0.9\textwidth]{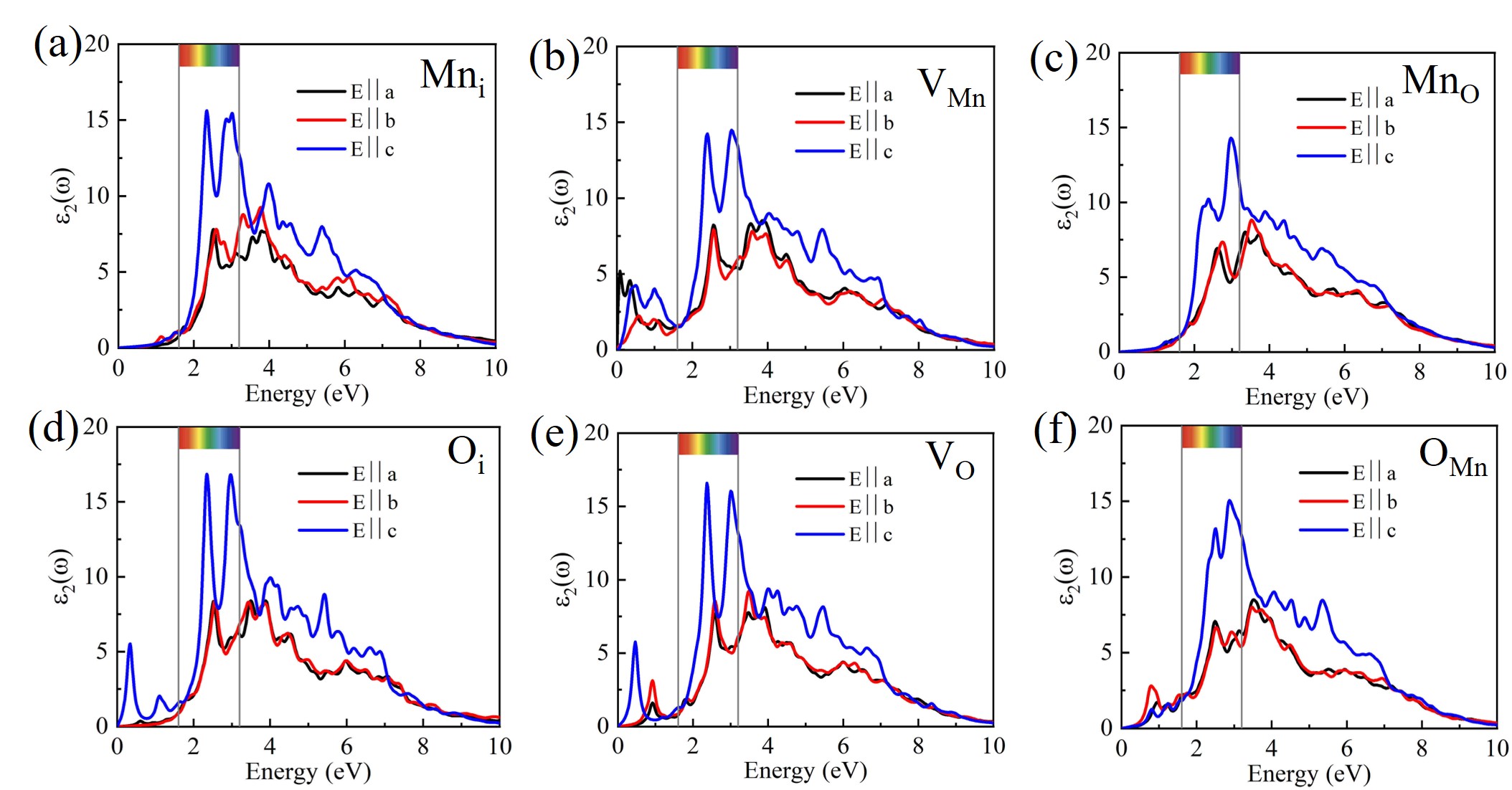}
\caption{(a)-(f) Imaginary part of dielectric constant $\varepsilon_2(\omega)$ for (a)-(f) for the isolated native defects in \(1 \times 1 \times 3\) $\alpha$-MnO$_2$ supercell, calculated using GW+BSE level of theory. Peaks indicate inter-band and defect-related transitions.}
\label{fig12}
\end{figure*}

\begin{figure*}[ht]
\centering
\includegraphics[width=0.9\textwidth]{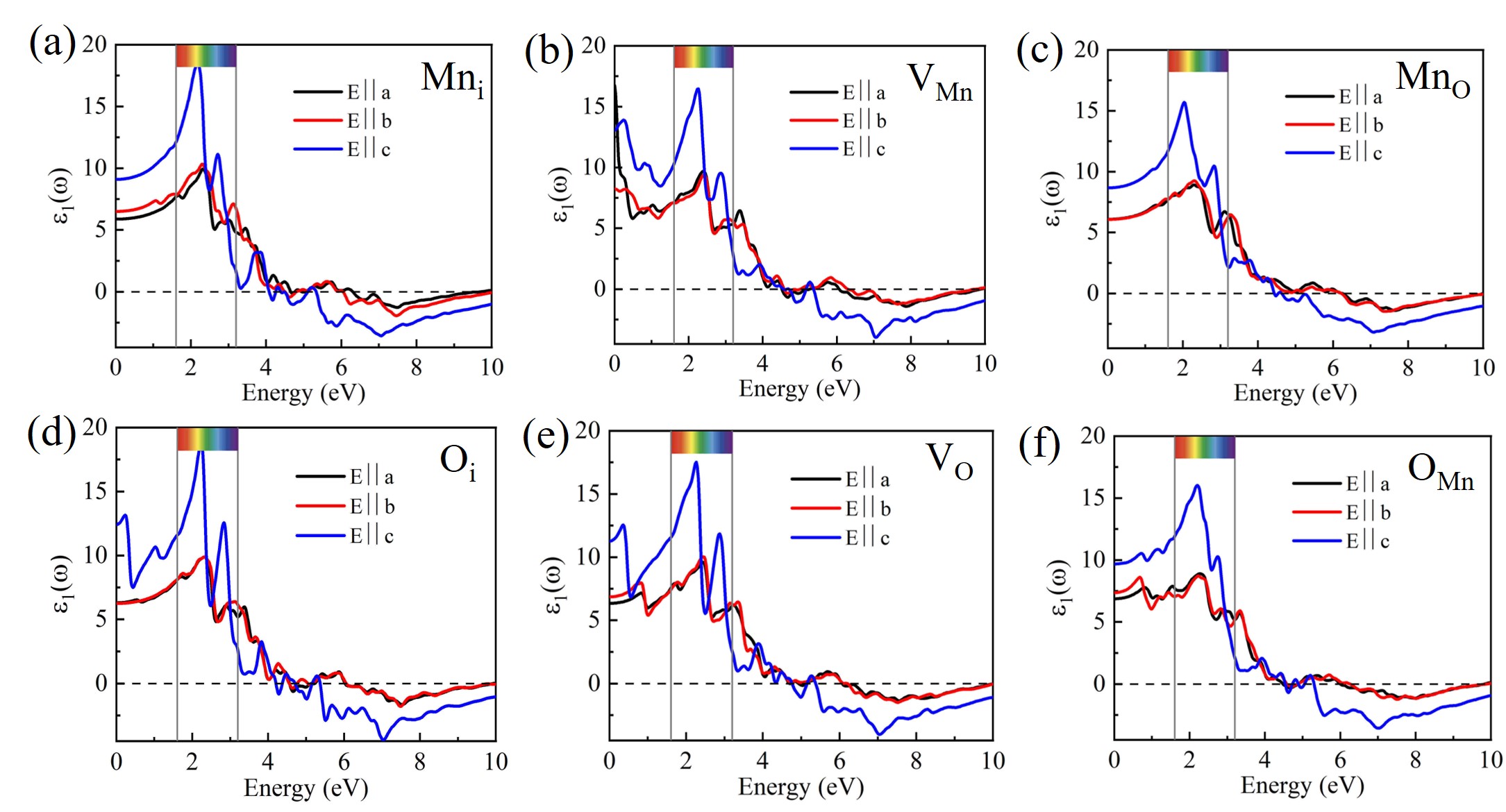}
\caption{Real part of dielectric constant $\varepsilon_1(\omega)$ for (a)-(f) isolated native defects in \(1 \times 1 \times 3\) $\alpha$-MnO$_2$ supercell, calculated using GW+BSE level of theory.}
\label{fig13}
\end{figure*}

\subsection{Optical Properties}

\begin{table*}[ht]
\caption{\label{table5}
Static dielectric constant ($\varepsilon$), and energy positions of the first peak (FP) and most intense (MI) peaks in $\varepsilon_2(\omega)$ for the stoichiometric \(1 \times 1 \times 3\) $\alpha$-MnO$_2$ supercell and various point defects along $E \parallel a$, $E \parallel b$, and $E \parallel c$, calculated at the GW+BSE level of theory. Energies are in eV.}
\begin{ruledtabular}
\begin{tabular}{lccc ccc ccc}
 & \multicolumn{3}{c}{$E \parallel a$} & \multicolumn{3}{c}{$E \parallel b$} & \multicolumn{3}{c}{$E \parallel c$} \\
Defect & $\varepsilon$ & FP & MI & $\varepsilon$ & FP & MI & $\varepsilon$ & FP & MI \\
\hline
Pristine            & 6.21 & 1.76 & 2.45 & 6.21 & 1.76 & 2.45 & 9.28 & 2.33 & 2.33, 2.88 \\
Mn$_\mathrm{i}$     & 5.88 & 1.17 & 2.51 & 6.51 & 1.13 & 3.77 & 9.10 & 1.23 & 2.34, 2.92 \\
O$_\mathrm{i}$      & 6.31 & 0.531 & 2.52 & 6.26 & 0.398 & 3.52 & 12.4 & 0.233 & 2.35, 2.91 \\
V$_\mathrm{Mn}$     & 16.7 & 0.081 & 2.57 & 8.25 & 0.592 & 2.57 & 13.1 & 0.511 & 2.35, 2.92 \\
V$_\mathrm{O}$      & 6.36 & 0.919 & 2.58 & 6.87 & 0.919 & 2.58 & 11.3 & 0.458 & 2.35, 2.91 \\
Mn$_\mathrm{O}$     & 6.10 & 1.29 & 3.34 & 6.10 & 1.23 & 3.51 & 8.67 & 1.24 & 2.34, 2.92 \\
O$_\mathrm{Mn}$     & 6.88 & 0.947 & 3.53 & 7.38 & 0.798 & 3.45 & 9.69 & 0.798 & 2.36, 2.86 \\
\end{tabular}
\end{ruledtabular}
\end{table*}

Previous sections demonstrate that intrinsic point defects significantly modify the electronic structure of $\alpha$-MnO$_2$. Since the optical response is directly governed by the underlying electronic states, we now investigate the defect-induced changes in the excitonic optical properties of $\alpha$-MnO$_2$ within the GW+BSE framework to assess its potential for electronic and optoelectronic applications.

The calculated BSE and independent-particle approximation (IPA) spectra of pristine $1 \times 1 \times 3$ $\alpha$-MnO$_2$ along the three crystallographic directions are shown in Fig.~\ref{fig11}. The optical response exhibits pronounced anisotropy between the in-plane ($E \parallel a,b$) and out-of-plane ($E \parallel c$) directions due to the tetragonal crystal structure. Inclusion of electron–hole interactions within the BSE framework enhances the absorption intensity and produces sharp structures, particularly for light polarized along the $c$ axis. The absorptive part of the dielectric function $\varepsilon_2(\omega)$ reveals an absorption onset at $\sim$1.6 eV in the BSE spectrum, which is absent in the IPA results owing to excitonic effects [cf. Figs.~\ref{fig11}(a) and \ref{fig11}(b)]. Although the overall spectral shape is redshifted relative to the IPA spectrum, the difference in the absorption onset is modest ($\sim$110 meV).

The first optically active excitonic peak appears at 1.76 eV for $E \parallel a,b$ and at 2.33 eV for $E \parallel c$, indicating a direction-dependent onset of optical transitions. This agrees well with experimental spectra reported for a structurally related variant of $\alpha$-MnO$_2$~\cite{gangwar2021structural}, as measurements for the pristine bulk phase are not available. The most intense peak occurs at 2.45 eV for the in-plane direction, whereas the spectrum for $E \parallel c$ exhibits a double-peaked structure with strong features at 2.33 and 2.88 eV, followed by a narrow peak at 3.19 eV and a shoulder around $\sim$3.4 eV. These results indicate that $\alpha$-MnO$_2$ is optically active in the visible and ultraviolet energy ranges and exhibits moderate anisotropy.

The dispersive part of the dielectric function $\varepsilon_1(\omega)$ is shown in Figs.~\ref{fig11}(c) and \ref{fig11}(d). The main spectral features occur at photon energies corresponding to the peaks in $\varepsilon_2(\omega)$, namely at 2.45 eV for $E \parallel a,b$ and 2.33 eV for $E \parallel c$. The static dielectric constant increases from 6.21 in the $(ab)$ plane to 9.28 along the $c$ axis, indicating stronger electronic polarizability along this direction. The calculated static dielectric constants, together with the energies of the first peak (FP) and the most intense (MI) peaks in $\varepsilon_2(\omega)$, are summarized in Table~\ref{table5}.

The introduction of intrinsic defects significantly modifies the low-energy optical response and dielectric screening, as shown in Figs.~\ref{fig12} and \ref{fig13}. The most prominent feature of the defective systems is the appearance of absorption peaks below 2 eV, whereas the higher-energy features in the visible region remain largely unchanged. Optical anisotropy persists in the defective structures and becomes more pronounced in the low-energy region of the spectra. For donor-type defects such as Mn$_\mathrm{i}$ and Mn$_\mathrm{O}$, the first optical peaks exhibit relatively small redshifts and reduced intensity compared with other defects [cf. Figs.~\ref{fig12}(a) and \ref{fig12}(c)]. This behavior can be attributed to enhanced electronic screening arising from the excess electrons associated with these donor states, as shown in Figs.~\ref{fig5}(b) and \ref{fig5}(d). These peaks involve transitions from electrons in the defect-induced gap states to the conduction bands.

In contrast, acceptor-type defects such as V$_\mathrm{Mn}$, O$_\mathrm{Mn}$, O$_\mathrm{i}$, and V$_\mathrm{O}$ produce pronounced redshifts in the absorption spectra due to reduced electronic screening. For example, V$_\mathrm{Mn}$ defect exhibits a defect-induced absorption characterized by a double-peaked structure in the low-energy region, with the lowest optical transition energy of $\sim$81 meV for $E \parallel a$. These excitations arise from transitions between the defect states and conduction-band states at different energies. Similarly, O$_\mathrm{Mn}$ introduces additional low-energy absorption features. Interestingly, the spectra for the amphoteric defects O$_\mathrm{i}$ and V$_\mathrm{O}$ display sharp low-energy peaks with maxima at 0.233 eV and 0.458 eV, respectively, for $E \parallel c$. In particular, V$_\mathrm{O}$ produces sharp absorption peaks in the low-energy region along all three directions.

To ensure that these defect-induced features are not artifacts of the relatively small supercell used in the GW+BSE calculations, we perform additional DFT+$U$ calculations for a larger $2 \times 2 \times 7$ supercell containing 672 atoms. While minor quantitative differences are observed, the positions of the main spectral features and the qualitative trends associated with different defects remain essentially unchanged [cf. Fig.~S15].

The changes in the low-energy absorption region are also reflected in the dielectric response shown in Fig.~\ref{fig13}. For donor-type defects, the static dielectric constant decreases slightly, whereas for acceptor-type defects it increases, particularly along the $c$ axis. Notably, V$_\mathrm{O}$ and O$_\mathrm{i}$ lead to enhanced dielectric constants for $E \parallel c$, while V$_\mathrm{Mn}$ produces a large dielectric response ($\varepsilon = 16.7$), indicating strong polarization effects. Overall, the optical spectra exhibit stronger intensity for light polarized along the $c$ direction, indicating that the dominant optical transitions occur along the tunnel direction of the crystal structure.

These results highlight the strong sensitivity of the optical response of $\alpha$-MnO$_2$ to the nature of intrinsic point defects. The introduction of defect states within the band gap enables optical transitions in the near-infrared and visible regions. Such defect-induced redshifts in absorption, particularly for vacancy defects, suggest potential routes for tuning the optical properties of $\alpha$-MnO$_2$ through controlled defect engineering for near-infrared optoelectronic applications.

\section{Conclusions}

In summary, we have carried out a detailed first-principles study of native point defects in C2-type antiferromagnetic $\alpha$-MnO$_2$, focusing on their thermodynamic stability, charge transition levels, electronic and optical properties. Mn$_\text{i}$ acts as shallow double donor, whereas V$_\text{O}$, O$_\text{i}$, and O$_\text{Mn}$ introduce deep states. Mn$_\text{O}$ exhibits both shallow and deep donor levels while V$_\text{Mn}$ gives rise to both shallow and deep acceptor levels. V$_\text{O}$, the most common defect in $\alpha$-MnO$_2$, emerges as an amphoteric compensating defect which differs from its conventional donor character in transition metal oxides. The calculated defect formation energies reveal intrinsic defect compensation arising from competing donor- and acceptor-type native defects, whose relative stability depends sensitively on the chemical environment. Notably, V$_\text{Mn}$ remains ionized throughout the band gap and behaves as a shallow acceptor, suggesting that it may become relevant under suitable non-equilibrium growth conditions. Overall, the excitonic optical response of $\alpha$-MnO$_2$ is highly sensitive to native defects, which induce defect-assisted low-energy absorption and increased dielectric screening without significantly altering the intrinsic high-energy O-2$p$ and Mn-3$d$ transitions. The present work is limited to the defect thermodynamics directly accessible from the calculated defect formation energies and charge transition levels. Nevertheless, our results establish a comprehensive understanding of the intrinsic defect physics and optical activity of $\alpha$-MnO$_2$, providing a strong foundation for future studies of equilibrium defect thermodynamics, defect engineering, and extrinsic doping for advanced electronic and optoelectronic applications.

\begin{acknowledgments}
High Performance Computing facilities of Indian Institute of Technology Bombay (IITB) and Bhabha Atomic Research Centre (BARC) were used in this work. AS thanks Prof. Alok Shukla for his support and encouragement and acknowledges IITB for the Institute Postdoctoral Fellowship (IPDF). 
\end{acknowledgments}

\bibliographystyle{unsrt}
\bibliography{references}

\end{document}


\title{Supplemental Material for \q{Anomalous behavior of native point defects in C2-ordered antiferromagnet $\alpha$-MnO$_2$}}

\author{Archana Sharma}
\affiliation{Department of Physics, Indian Institute of Technology Bombay, Mumbai 400076, India}
\author{Brahmananda Chakraborty}
\email{brahma@barc.gov.in}

\affiliation{High Pressure and Synchrotron Radiation Physics Division, Bhabha Atomic Research Centre, Trombay, Mumbai 400085, India}
\altaffiliation[Also at ]{Homi Bhabha National Institute, Mumbai 400085, India}

\maketitle

This supporting document contains the following entries: \vspace{-50mm}

\bigskip

\vspace{100mm}

\bigskip

\begin{tabular}{ll}
S1. & Benchmarking of the Hubbard parameter $U$ ................................................................. S2 \\
S2. & SOC included band structure of pristine $\alpha$-MnO$_2$ ........................................................ S3 \\
S3. & Electronic structures and optimized geometries of native defects ............................ S4-S5 \\
S4. & Charge-state dependence of finite-size corrections ................................................... S5-S6 \\
S5. & Defect formation energies and charge transition levels ............................................ S7-S13 \\
S6. & Effect of the Hubbard $U$ on charge transition levels of V$_\text{O}$.......................................... S14\\
S7. & Comparison between FNV and eFNV stability diagrams ............................................ S15 \\
S8. & Supercell-size convergence of defect-induced optical spectra ....................................... S16
\end{tabular}

\newpage

\subsection*{S1. Benchmarking of the Hubbard parameter $U$}

\begin{figure*}[ht]
\includegraphics[width=0.9\linewidth]{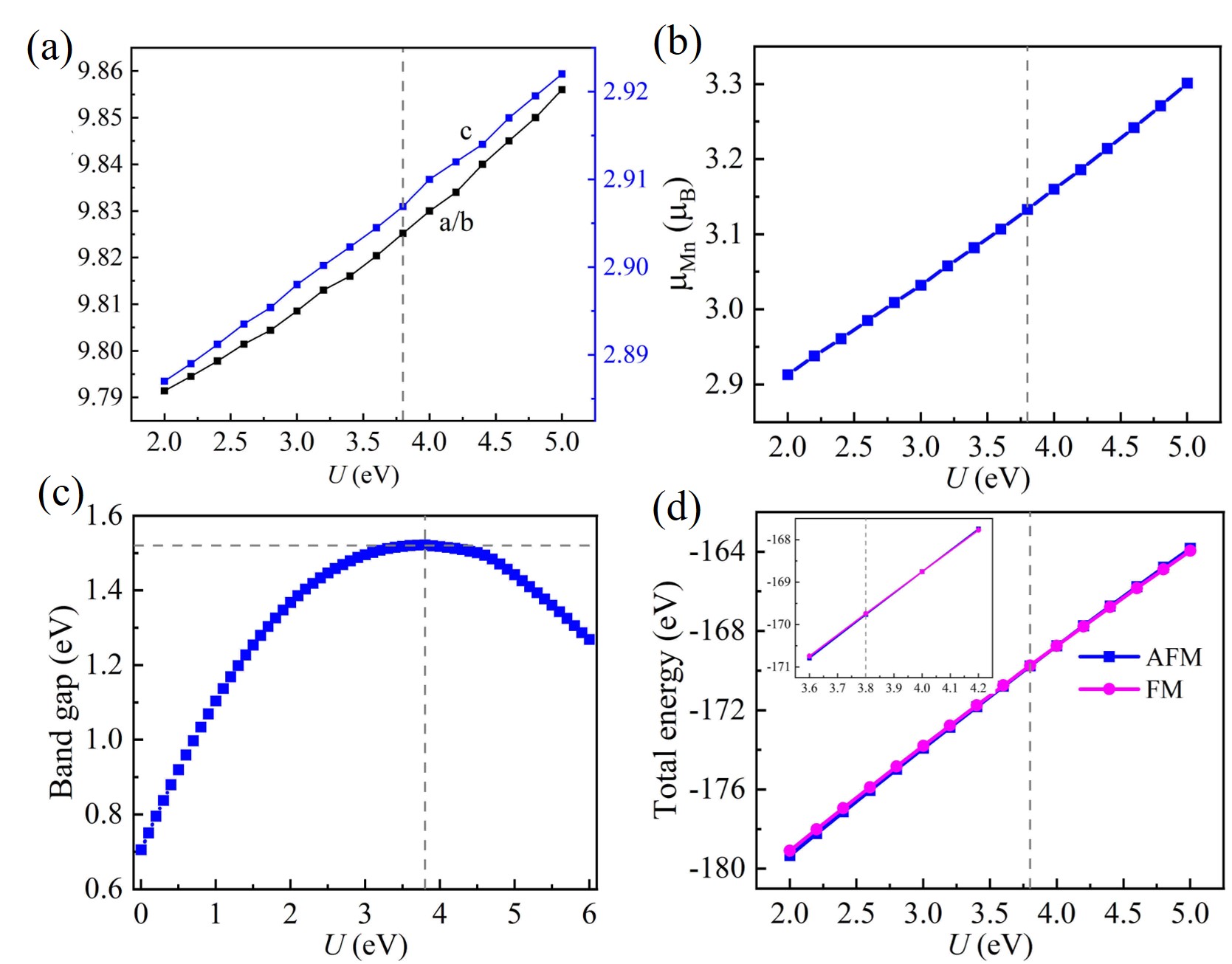}\caption{Benchmarking of Hubbard parameter ($U$): Variation of U with (a) lattice parameters ($a$, $b$ and $c$), local magnetic moment on Mn ($\mu_\text{Mn}$), band gap (eV), and total energies of $\alpha$-MnO$_2$ with AFM and FM configurations.}
\label{figS1}
\end{figure*}

The Hubbard parameter ($U$) employed for Mn $3d$ states is benchmarked by comparing the structural, magnetic, and electronic properties of $\alpha$-MnO$_2$ with available experimental data. Fig.~\ref{figS1} summarizes the variation of lattice parameters, local magnetic moment, band gap, and relative energies of the C2-ordered antiferromagnetic (AFM) and ferromagnetic (FM) configurations as a function of $U$. Based on these results, $U=3.8$ eV is adopted throughout this work.

\newpage

\subsection*{S2. SOC included band structure of pristine $\alpha$-MnO$_2$}

\begin{figure*}[ht]
\includegraphics[width=0.5\linewidth]{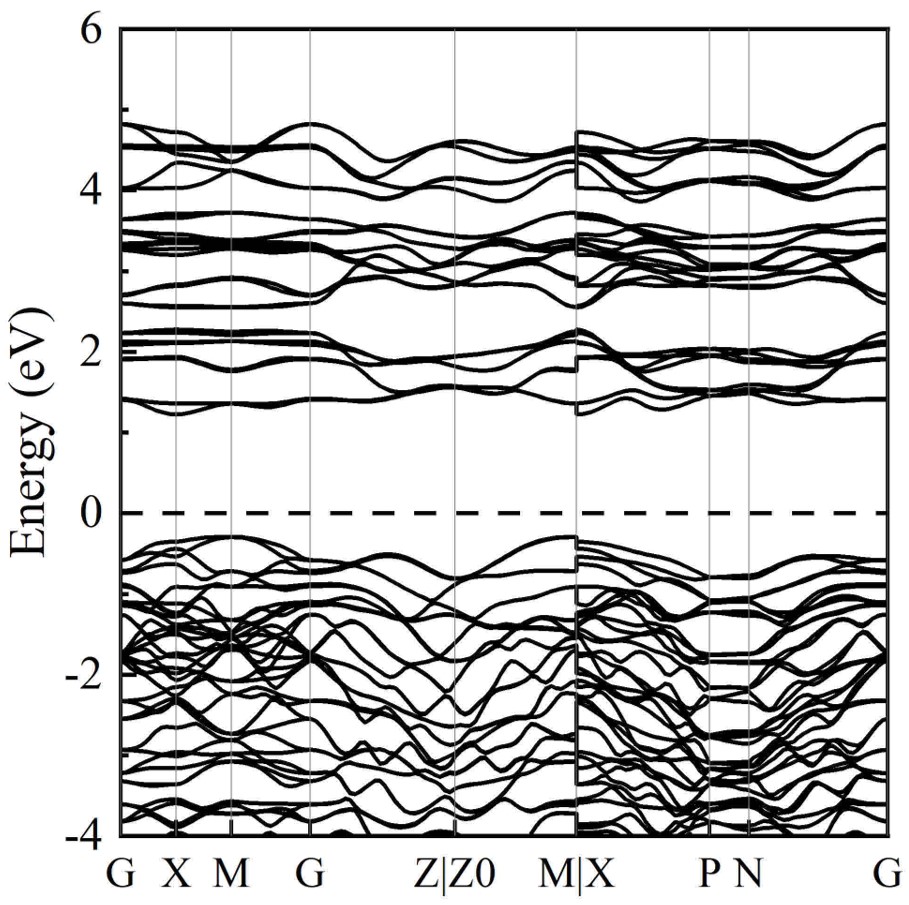}\caption{Electronic band structure of primitive unit cell of $\alpha$-MnO$_2$ including SOC effect.}
\label{figS2}
\end{figure*}

To assess the influence of spin-orbit coupling (SOC), the electronic band structure of pristine $\alpha$-MnO$_2$ was calculated with SOC included. As shown in Fig.~\ref{figS2}, SOC has only a negligible effect near the band edges.

\newpage

\subsection*{S3. Electronic structures and optimized geometries of native defects}

\begin{figure*}[ht]
\includegraphics[width=0.9\linewidth]{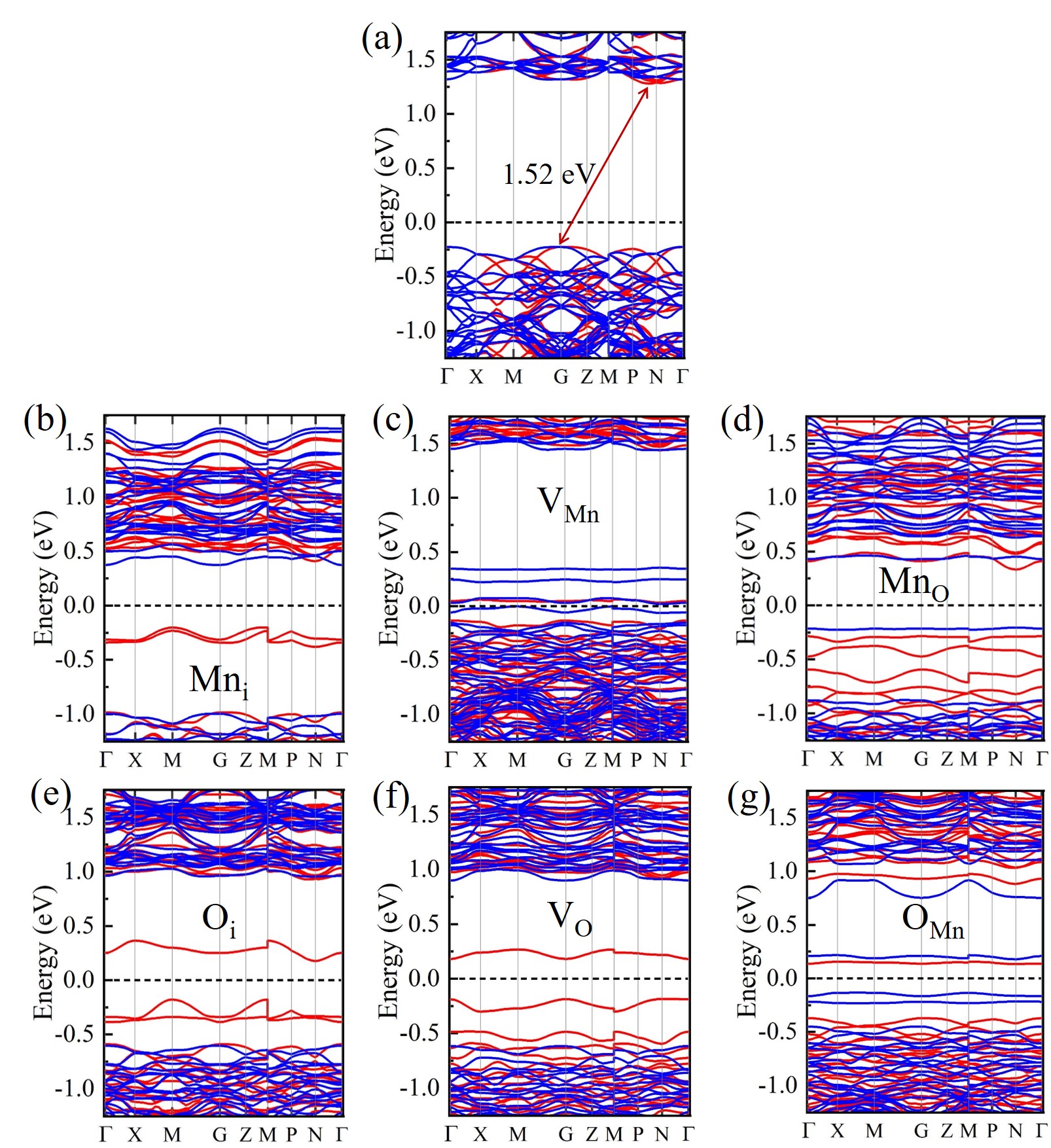}\caption{Spin-polarized electronic band structure of (a) \(1 \times 1 \times 3\) $\alpha$-MnO$_2$ supercell with (b, e) interstitials (Mn$_\text{i}$, O$_\text{i}$), (c, f) vacancies (V$_\text{Mn}$, V$_\text{O}$), and (d, g) antisites (Mn$_\text{O}$, O$_\text{Mn}$), calculated using DFT+$U$ level of theory. Red(blue) color denotes spin-up(spin-down) bands. The Fermi level is scaled to 0 eV. High symmetry points Z\textbar Z0 and M\textbar X are denoted by Z and M, for brevity.}
\label{figS3} 
\end{figure*}

\begin{figure*}[ht]
\includegraphics[width=0.9\linewidth]{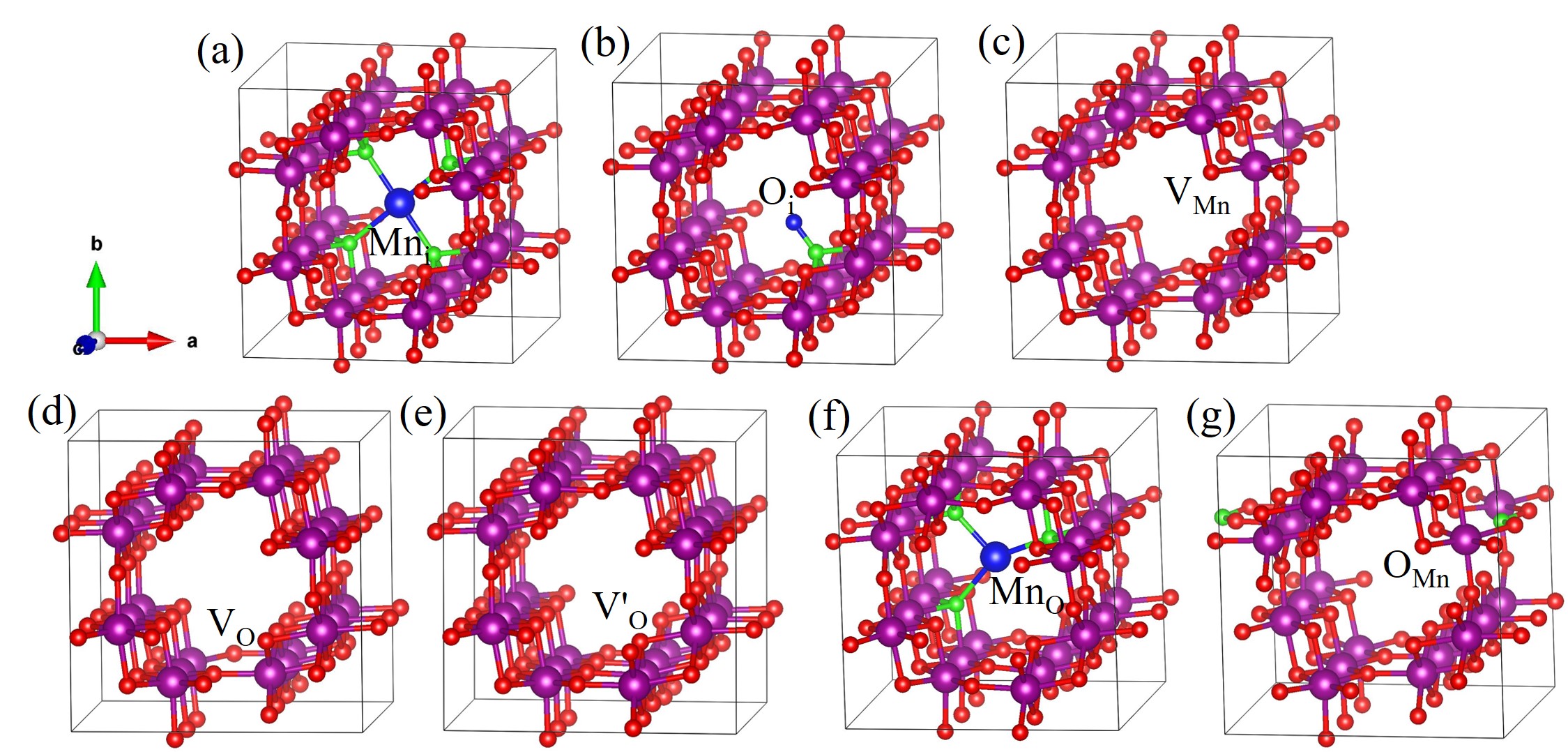}\caption{Optimized defect geometries of \(1 \times 1 \times 3\) $\alpha$-MnO$_2$ supercell with (a, b) interstitials (Mn$_\text{i}$, O$_\text{i}$), (c-e) vacancies (V$_\text{Mn}$, V$_\text{O}$, V$_\text{O}$'), and (f,g) antisites (Mn$_\text{O}$, O$_\text{Mn}$). Interstitial atoms (Mn, O) and Mn$_\text{O}$ are shown in blue while neighboring O atoms in green. O$_\text{Mn}$ is depicted in green color.}
\label{figS4} 
\end{figure*}

Fig.~\ref{figS3}(a) shows the spin-polarized electronic band structure of pristine 1 $\times$ 1 $\times$ 3 $\alpha$-MnO$_2$. Owing to the enlarged supercell, pronounced band-folding effects are observed, leading to a higher density of electronic bands within the Brillouin zone. Figs.~\ref{figS3}(b)–\ref{figS3}(g) present the spin-polarized band structures of 1 $\times$ 1 $\times$ 3 $\alpha$-MnO$_2$ supercell containing the various native point defects, highlighting the defect-induced electronic states that emerge within the band gap. The corresponding fully relaxed atomic structures for all defect configurations are shown in Fig.~\ref{figS4}.

\subsection*{S4. Charge-state dependence of finite-size corrections}

\begin{figure*}[ht]
\includegraphics[width=0.7\linewidth]{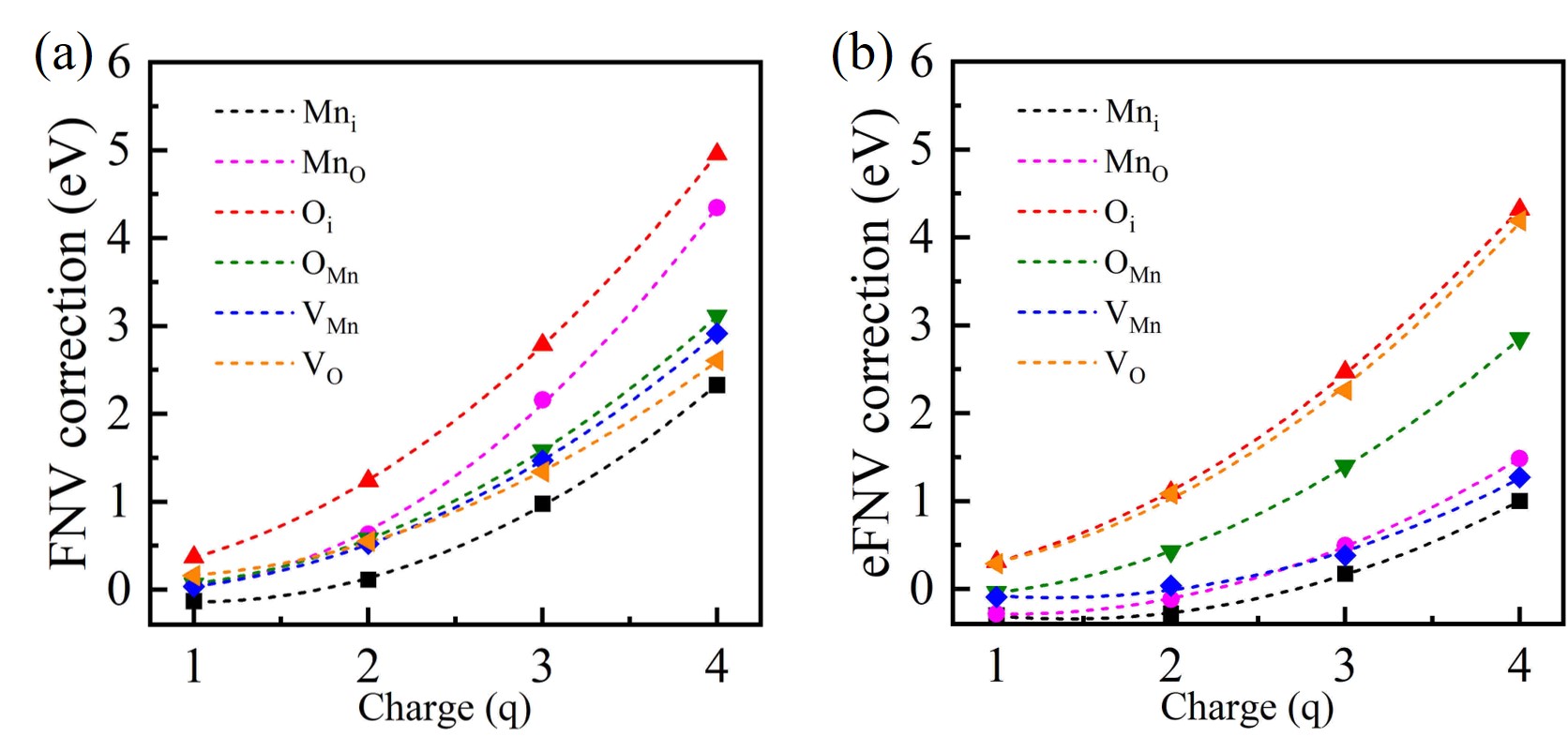}\caption{(a) FNV scaling and (b) extended-FNV scaling of the charged defects in $\alpha$-MnO$_2$ supercell with respect to various charges (q). For the defects O$_\text{i}$,  O$_\text{Mn}$,  V$_\text{Mn}$ and  V$_\text{O}$, q takes the negative values. The correction energies are fitted with a function of aq$^{2}$ + bq + c.}
\label{figS5}
\end{figure*}

Finite-size corrections for charged defects were evaluated using both the Freysoldt–Neugebauer–Van de Walle (FNV) and extended-FNV (eFNV) schemes. As shown in Fig. \ref{figS5}, the correction exhibits an overall quadratic dependence on the charge state but is more accurately described by (E$_{\text{corr}}$ = aq$^{2}$ + bq + c), where the linear term originates from potential alignment. This contribution is particularly important for low charge states, leading to deviations from simple (q$^2$) scaling.

\begin{table}[ht]
\caption{\label{tableS1}
Electrostatic correction terms ($E_\text{corr}^\text{FNV/eFNV}$in eV) for each defect type in \(1 \times 1 \times 3\) $\alpha$-MnO$_2$ supercell for stable charge states: lattice correction term ($E_{lat}$) and  potential alignment term ($\Delta V_\text{q}^\text{iso}$) for FNV corrections, and point-charge correction term ($E_\text{PC}$) and point-charge potential alignment term ($\Delta V_\text{q}^\text{aniso}$) for extended-FNV corrections.}
\begin{ruledtabular}
\begin{tabular}{lccccc}
\textbf{Defect type} & \textbf{Charge state} & \textbf{$E_\text{lat}$} (\textbf{$E_\text{PC}$}) & \textbf{$\Delta V_\text{q}^\text{iso}$} (\textbf{$\Delta V_\text{q}^\text{aniso}$}) & \textbf{$E_\text{corr}^\text{FNV}$} & \textbf{$E_\text{corr}^\text{eFNV}$} \\
\hline
Mn$_\text{i}$   & $1+$   & 0.238(0.131)  & 0.369(0.422) & $-0.130$  & $-0.298$ \\
                & $2+$   & 0.954(0.523)  & 0.839(0.802) & 0.115  & $-0.279$ \\
\hline
O$_\text{i}$    & $1+$   & 0.245(0.131)  & 0.088(0.005)    & 0.157   & 0.126  \\
                & $1-$   & 0.245(0.131)  & $-0.125$($-0.181$) & 0.371   & 0.312  \\
\hline
V$_\text{Mn}$   & $1-$ & 0.249(0.131) & 0.213(0.223)    & 0.037   & $-0.092$  \\
                & $2-$ & 0.997(0.523) & 0.474(0.484)    & 0.523   & 0.039  \\
                & $3-$ & 2.24(1.18)   & 0.776(0.793)    & 1.47    & 0.384  \\
\hline
V$_\text{O}$    & $1+$   & 0.162(0.131)  & $-0.022$($-0.074$) & 0.183 & 0.205  \\
                & $1-$   & 0.162(0.131)  & 0.001($-0.157$)    & 0.161 & 0.288  \\
\hline
Mn$_\text{O}$   & $1+$   & 0.238(0.131)  & 0.183(0.412) & 0.055  & $-0.281$  \\
                & $2+$   & 0.954(0.523)  & 0.323(0.641) & 0.631  & $-0.118$  \\
\hline
O$_\text{Mn}$   & $1-$ & 0.240(0.131) & 0.173(0.166)    & 0.067   & $-0.036$  \\
                & $2-$ & 0.961(0.523) & 0.380(0.095)    & 0.581   & 0.428  \\
\end{tabular}
\end{ruledtabular}
\end{table}

\clearpage

\subsection*{S5. Defect formation energies and charge transition levels}

\begin{figure*}[ht]
\includegraphics[width=0.9\linewidth]{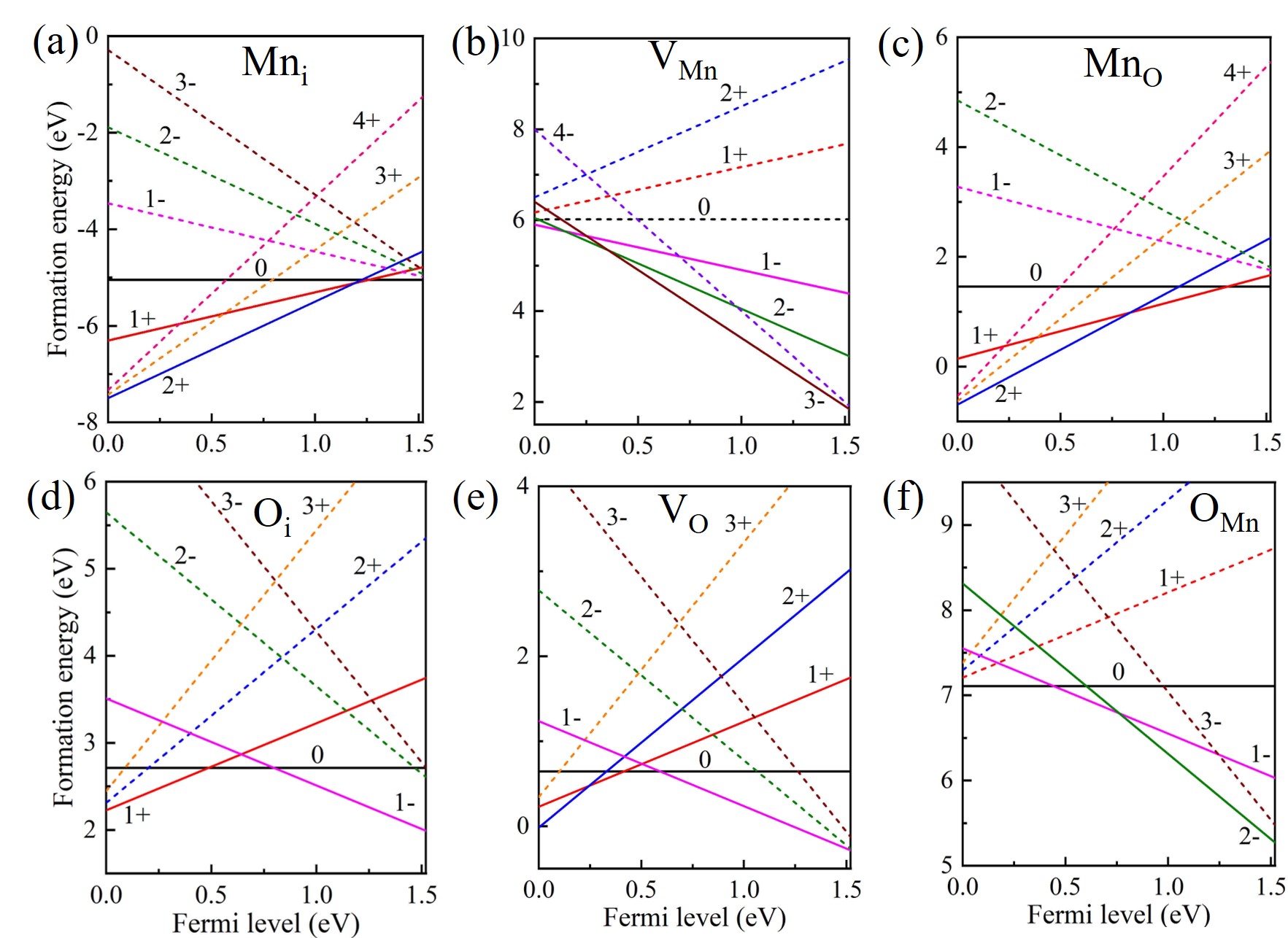}\caption{(a)-(f) Uncorrected formation energies of point defects in various charge states as a function of the Fermi level under Mn-rich conditions in \(1 \times 1 \times 3\) $\alpha$-MnO$_2$ supercell. The stable (unstable) charge states are shown by the solid (dashed) lines.}
\label{figS6} 
\end{figure*}

\begin{figure*}[ht]
\includegraphics[width=0.9\linewidth]{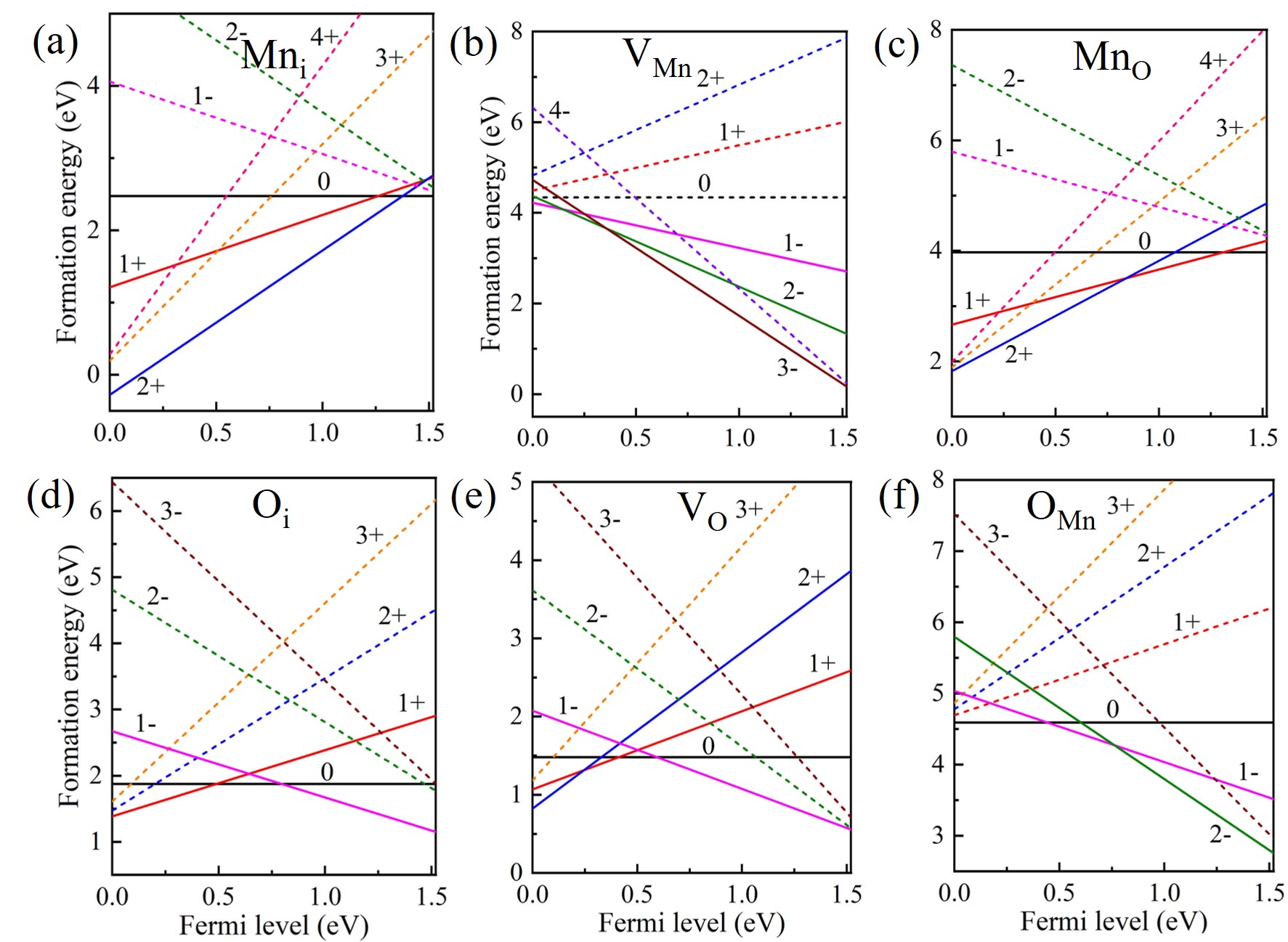}\caption{(a)-(f) Uncorrected formation energies of point defects in various charge states as a function of the Fermi level under O-rich conditions in \(1 \times 1 \times 3\) $\alpha$-MnO$_2$ supercell. The stable (unstable) charge states are shown by the solid (dashed) lines.}
\label{figS7} 
\end{figure*}

\begin{figure*}[ht]
    \centering
    \includegraphics[width=0.9\linewidth]{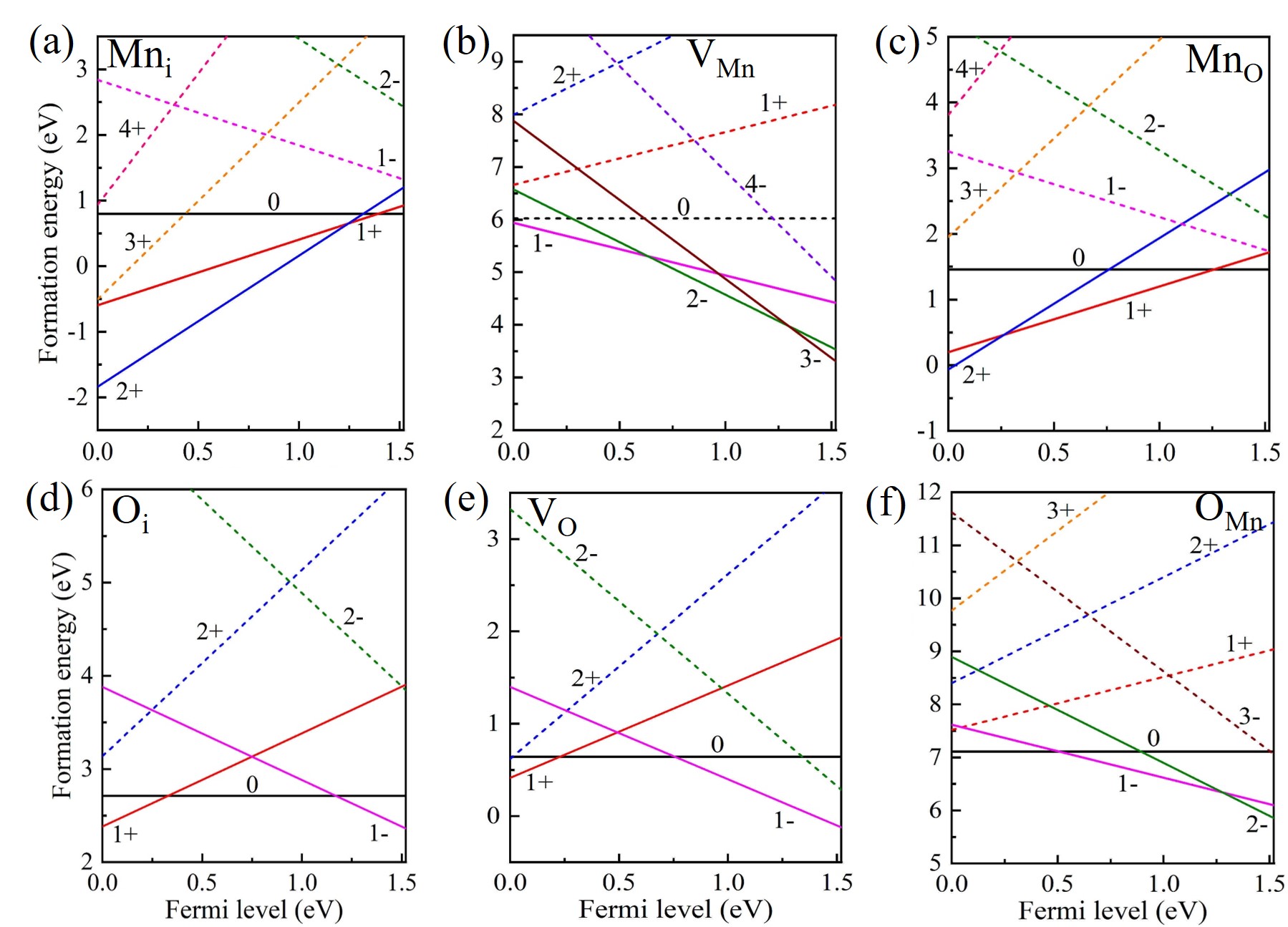}
    \caption{(a)–(f) FNV corrected formation energies of native point defects in various charge states under Mn-rich conditions in \(1 \times 1 \times 3\) $\alpha$-MnO$_2$ supercell, plotted as a function of the Fermi level. Solid (dashed) lines represent thermodynamically stable charge (metastable or unstable) states. The Fermi level spans the PBE+$U$ band gap, ranging from the VBM at 0 eV to the CBM at 1.52~eV.}
    \label{figS8}
\end{figure*}

\begin{figure*}[ht]
    \centering
    \includegraphics[width=0.9\linewidth]{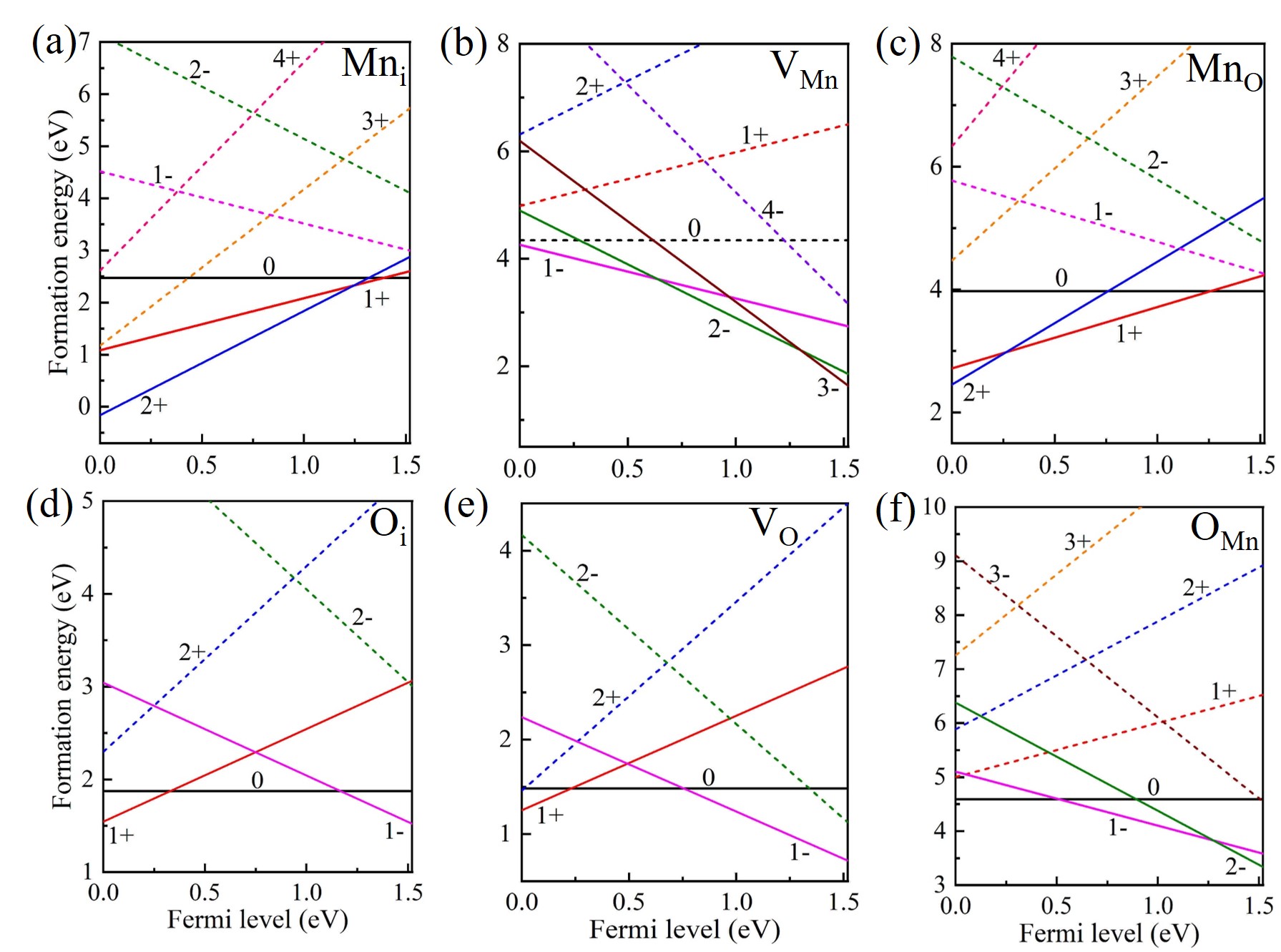}
    \caption{(a)–(f) FNV corrected formation energies of native point defects in various charge states under O-rich conditions in \(1 \times 1 \times 3\) $\alpha$-MnO$_2$ supercell, plotted as a function of the Fermi level. Solid (dashed) lines represent thermodynamically stable charge (metastable or unstable) states.}
    \label{figS9}
\end{figure*}

\begin{figure*}[ht]
    \centering
    \includegraphics[width=0.9\linewidth]{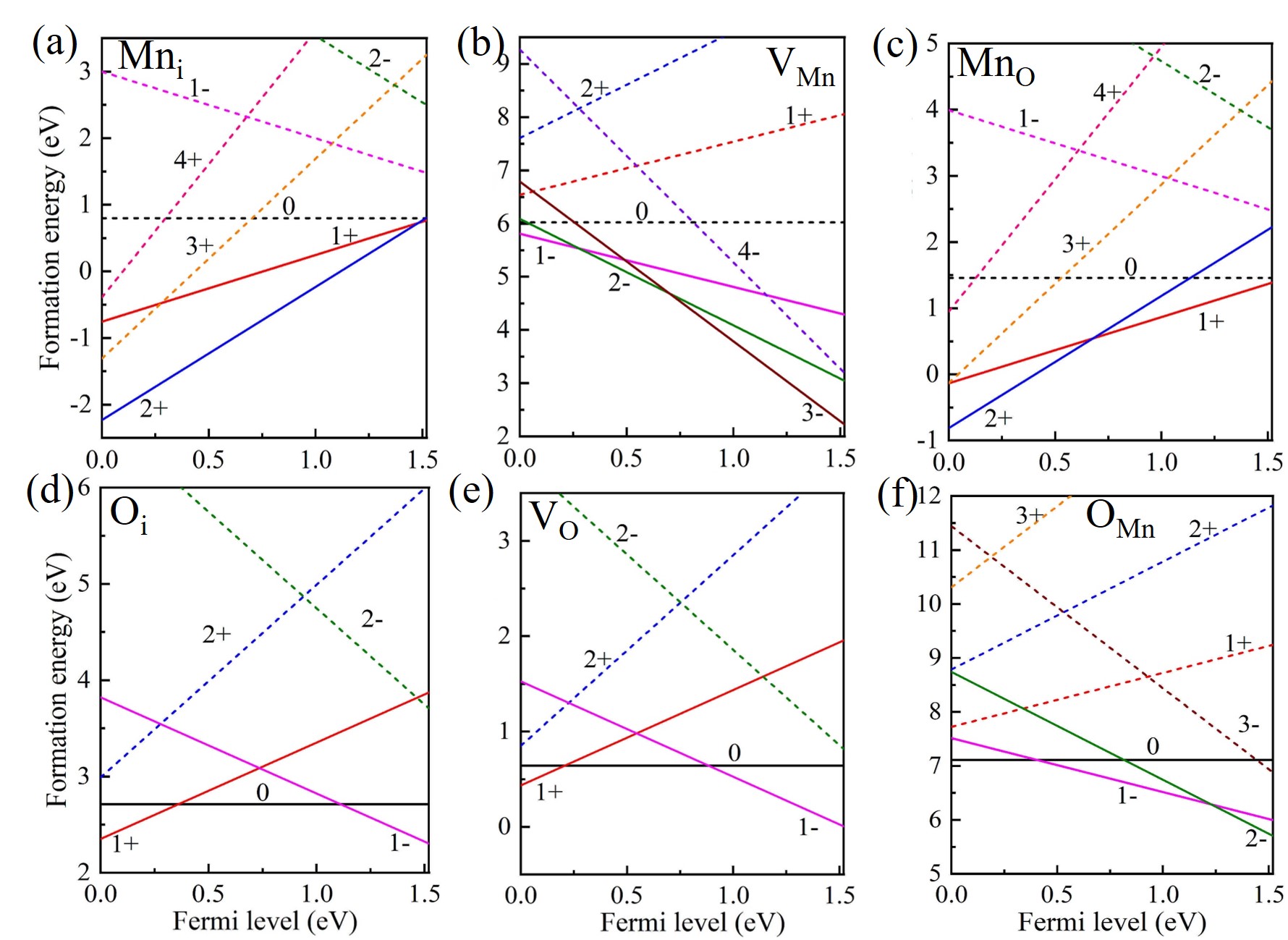}
    \caption{(a)–(f) Extended-FNV corrected formation energies of native point defects in various charge states under Mn-rich conditions in \(1 \times 1 \times 3\) $\alpha$-MnO$_2$ supercell, plotted as a function of the Fermi level. Solid (dashed) lines represent thermodynamically stable charge (metastable or unstable) states.}
    \label{figS10}
\end{figure*}

\begin{figure*}[ht]
    \centering
    \includegraphics[width=0.9\linewidth]{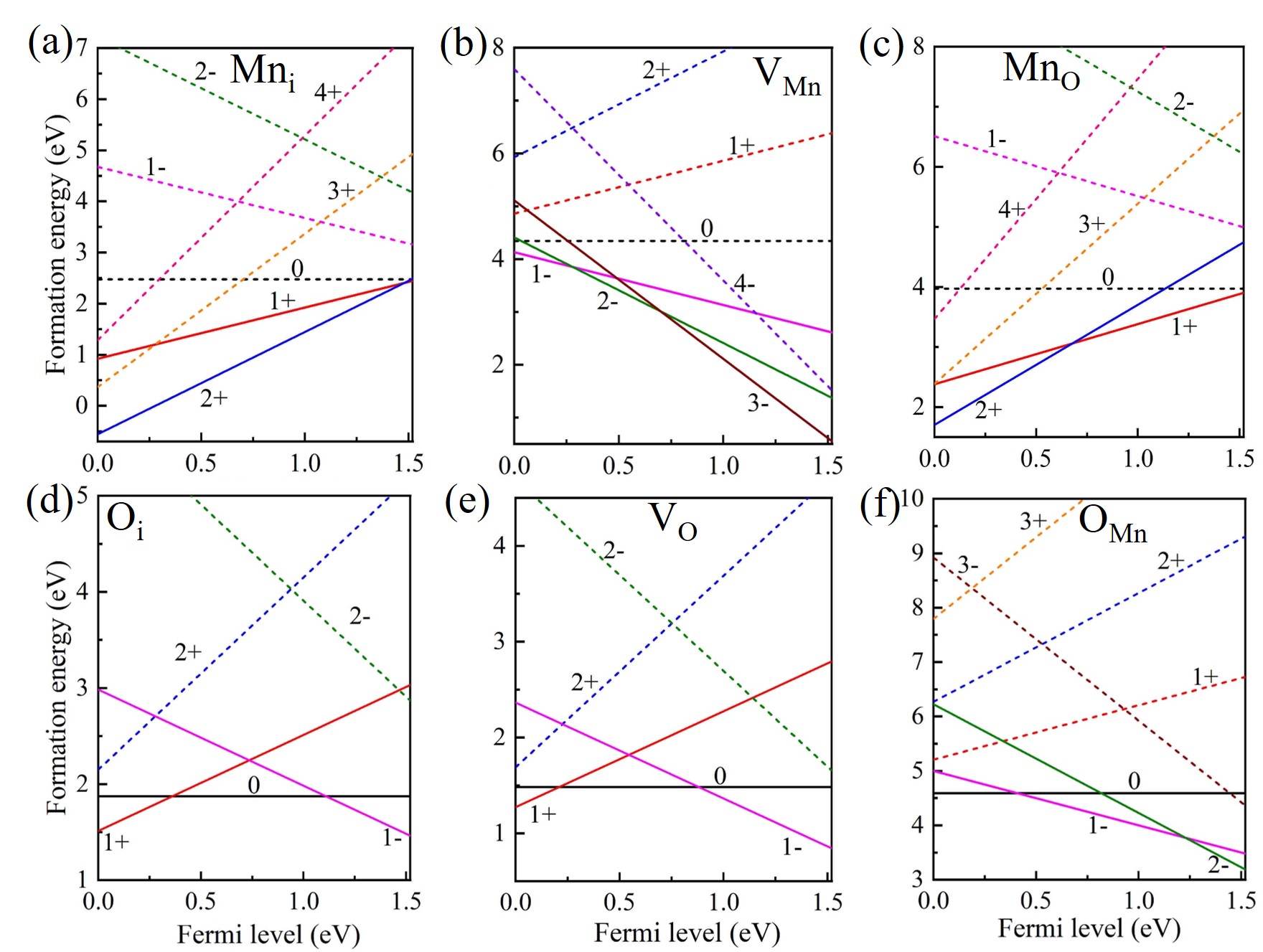}
    \caption{(a)–(f) Extended-FNV corrected formation energies of native point defects in various charge states under O-rich conditions in \(1 \times 1 \times 3\) $\alpha$-MnO$_2$ supercell, plotted as a function of the Fermi level. Solid (dashed) lines represent thermodynamically stable charge (metastable or unstable) states.}
    \label{figS11}
\end{figure*}

This section presents the defect formation energies obtained using uncorrected [Figs. \ref{figS6} and \ref{figS7}], FNV-corrected [Figs. \ref{figS8} and \ref{figS9}]and eFNV-corrected [Figs. \ref{figS10} and \ref{figS11}] calculations under Mn-rich and O-rich conditions. The corrected formation energies at the band edges are listed in Table~\ref{tableS2}, while the corresponding thermodynamic charge-transition levels are compiled in Table~\ref{tableS3}.

\begin{table}[ht]
\caption{\label{tableS2}
Corrected formation energies (in eV) of charged defects in the \(1 \times 1 \times 3\) \(\alpha\)-MnO$_2$ (\(2 \times 2 \times 7\)) supercell at band edges, for both Mn-rich and O-rich environments. Extended FNV corrections are applied to all the defects except V$_\mathrm{Mn}$ and V$_\mathrm{O}$ defects where standard FNV corrections are used.}
\centering
\begin{ruledtabular}
\begin{tabular}{lccccc}
\multirow{2}{*}{\textbf{Defect type}} & \multirow{2}{*}{\textbf{Charge state}} & \multicolumn{2}{c}{\textbf{Mn-rich}} & \multicolumn{2}{c}{\textbf{O-rich}} \\
\cline{3-6}
& & \textbf{VBM} & \textbf{CBM} & \textbf{VBM} & \textbf{CBM} \\
\hline
Mn$_\mathrm{i}$ & 1  & $-0.755$ ($-0.755$) & 0.764 ($0.764$) & 0.923 (0.922) & 2.44 (2.44) \\
                & 2  & $-2.23$ ($-2.23$) & 0.807 ($-0.808$) & $-0.555$ ($-0.554$) & 2.48 (2.48) \\
O$_\mathrm{i}$  & 1  & $2.35$ ($2.38$) & $3.87$ ($3.90$) & $1.51$ ($1.54$) & $3.03$ ($3.06$) \\
                & $-1$ & $3.82$ ($3.85$) & $2.30$ ($2.33$) & $2.98$ ($3.01$) & $1.46$ ($1.49$) \\
V$_\mathrm{Mn}$ & $-1$ & $5.94$ ($5.97$) & $4.42$ ($4.45$) & $4.26$ ($4.29$) & $2.74$ ($2.77$) \\
                & $-2$ & $6.57$ ($6.60$) & $3.53$ ($3.56$) & $4.89$ ($4.92$) & $1.85$ ($1.88$) \\
                & $-3$ & $7.87$ ($7.84$) & $3.31$ ($3.28$) & $6.19$ ($6.16$) & $1.63$ ($1.60$) \\
V$_\mathrm{O}$  & 1  & $0.414$ ($0.414$) & $1.93$ ($1.93$) & $1.25$ ($1.25$) & $2.77$ ($2.77$) \\
                & $-1$ & $1.40$ ($1.40$) & $-0.122$ ($-0.123$) & $2.24$ ($2.24$) & $0.717$ ($0.717$) \\
Mn$_\mathrm{O}$ & 1  & $-0.135$ ($-0.129$) & $1.38$ ($1.39$) & $2.38$ ($2.39$) & $3.90$ ($3.90$) \\
                & 2  & $-0.812$ ($-0.806$) & $2.23$ ($2.24$) & $1.70$ ($1.70$) & $4.74$ ($4.75$) \\
O$_\mathrm{Mn}$ & $-1$ & $7.51$ ($7.56$) & $5.99$ ($6.04$) & $4.99$ ($5.04$) & $3.48$ ($3.53$) \\
                & $-2$ & $8.74$ ($8.68$) & $5.70$ ($5.65$) & $6.22$ ($6.17$) & $3.18$ ($3.12$) \\
\end{tabular}
\end{ruledtabular}
\end{table}

\begin{table}[ht]
\centering
\caption{Charge transition levels (in eV) for native point defects in \(2 \times 2 \times 7\) $\alpha$-MnO$_2$ supercell, referenced to the VBM (corrected using FNV approach).}
\label{tableS3}
\setlength{\tabcolsep}{10pt}
\renewcommand{\arraystretch}{1.2}
\setlength{\tabcolsep}{3pt}  
\begin{tabular}{lccccc}
\hline\hline
\multirow{2}{*}{\shortstack{Defect\\type}} & \multicolumn{5}{c}{Charge transition levels} \\
\cline{2-6}
& (2+/1+) & (1+/0) & (0/1$-$) & (1$-$/2$-$) & (2$-$/3$-$) \\
\hline
Mn$_\text{i}$   & 1.24 & 1.39 &       &       &       \\
O$_\text{i}$    &       & 0.329 & 1.18 &       &       \\
V$_\text{Mn}$   &       &       &       & 0.659 & 1.33 \\
V$_\text{O}$    &       & 0.228 & 0.755 &       &       \\
Mn$_\text{O}$   & 0.284 & 1.25 &       &       &       \\
O$_\text{Mn}$   &       &       & 0.512 & 1.23 &       \\
\hline\hline
\end{tabular}
\end{table}

\clearpage

\subsection*{S6. Effect of the Hubbard $U$ on charge transition levels of V$_\text{O}$}

\begin{figure*}[ht]
\includegraphics[width=0.5\linewidth]{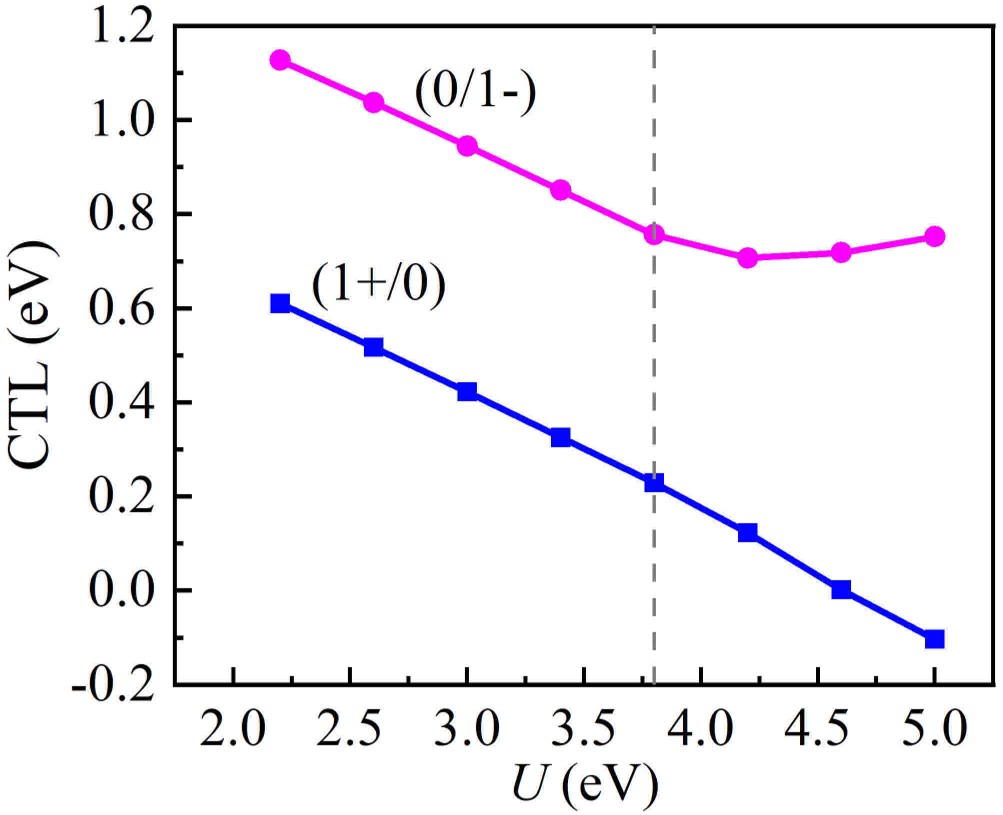}\caption{Variation of charge transition levels (CTLs) of V$_\text{O}$ as a function of Hubbard parameter $U$.}
\label{figS12}
\end{figure*}

Charge transition levels (CTLs) of V$_\text{O}$ calculated with Hubbard (U) values ranging from 2 to 5 eV are shown in Fig.~\ref{figS12}. Increasing $U$ enhances the localization of Mn-3$d$ states, causing the donor level to shift toward the VBM, while the acceptor level moves toward midgap up to $U$ = 3.8 eV before shifting back toward the CBM. Within the physically relevant range $U$ $\leq$ 3.8) eV, both donor and acceptor CTLs remain inside the band gap, reflecting the robustness of the amphoteric nature of V$_\text{O}$ against the choice of $U$.

\clearpage

\subsection*{S7. Comparison between FNV and eFNV stability diagrams}

\begin{figure*}[ht]
\includegraphics[width=0.7\linewidth]{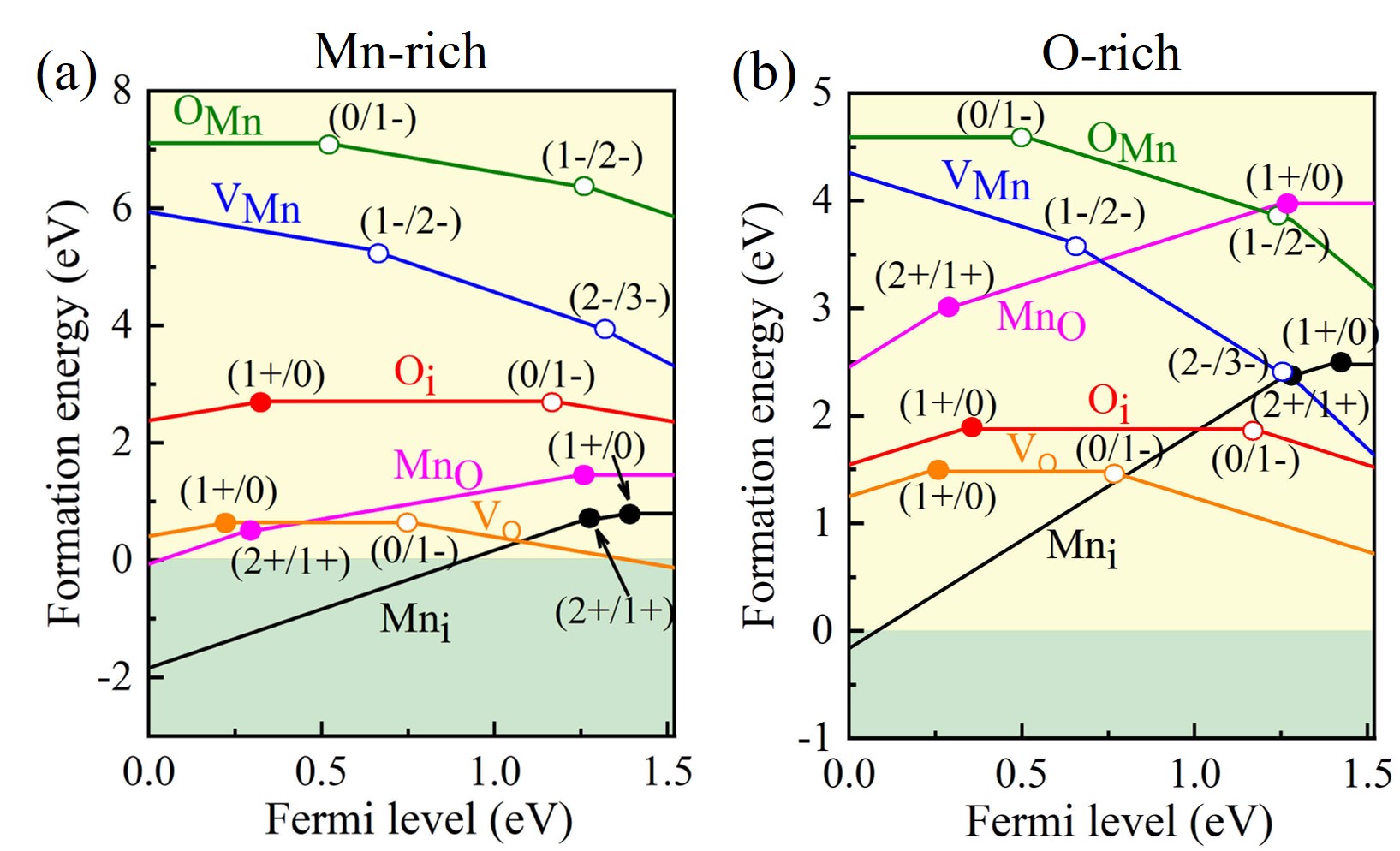}\caption{(a), (b) Variation of formation energies (FNV corrected) for stable charge states of native point defects in \(1 \times 1 \times 3\) $\alpha$-MnO$_2$ supercell as a function of the Fermi level under Mn-rich and O-rich conditions. Charge transition levels are indicated in parentheses. The green (yellow) shaded region represents the regime of spontaneous (non-spontaneous) defect formation.}
\label{figS13} 
\end{figure*}

\begin{figure*}[ht]
\includegraphics[width=0.7\linewidth]{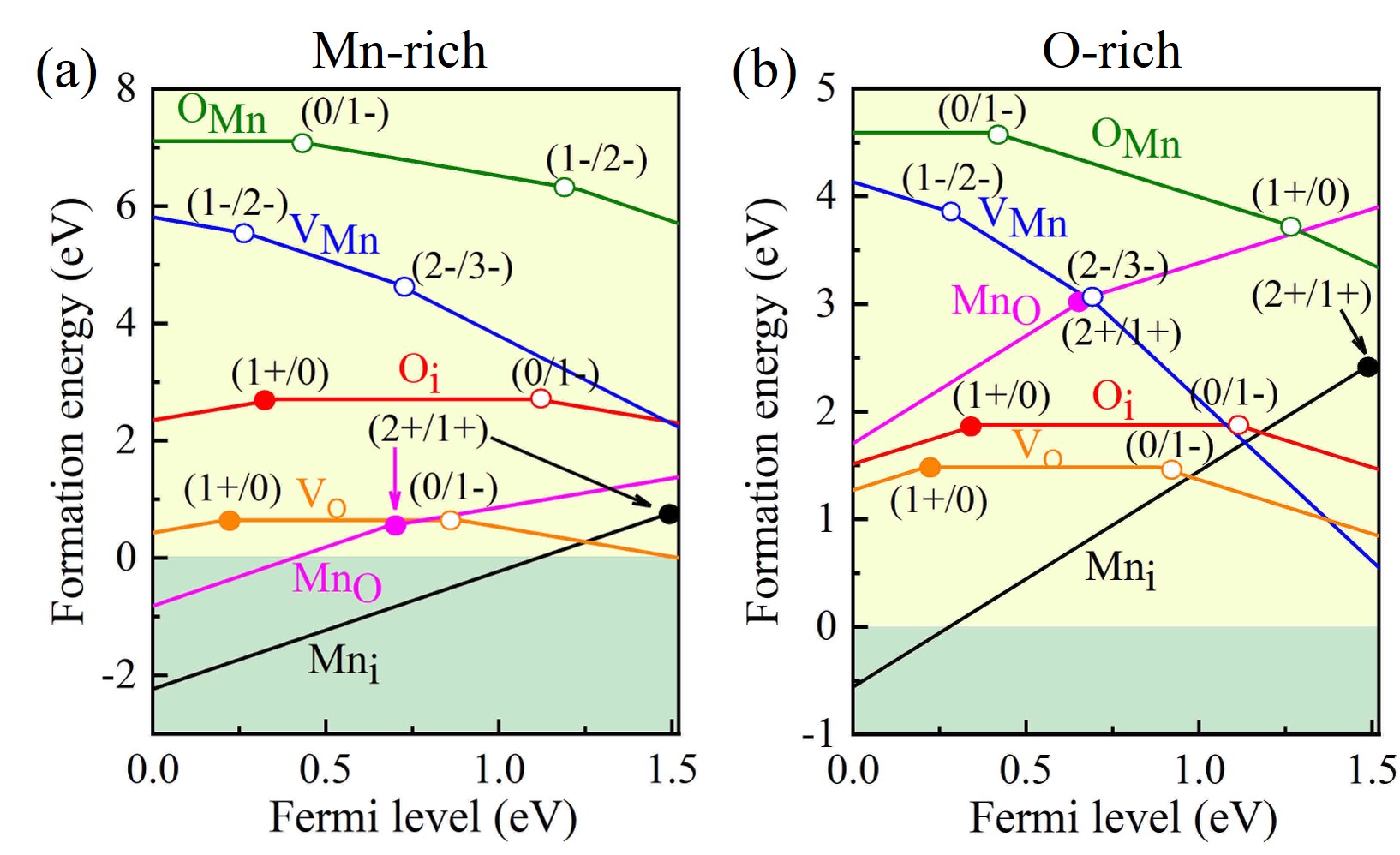}\caption{(a), (b) Variation of formation energies (eFNV corrected) for stable charge states of native point defects in \(1 \times 1 \times 3\) $\alpha$-MnO$_2$ supercell as a function of the Fermi level under Mn-rich and O-rich conditions. Charge transition levels are indicated in parentheses. The green (yellow) shaded region represents the regime of spontaneous (non-spontaneous) defect formation.}
\label{figS14} 
\end{figure*}

To examine the sensitivity of the defect stability analysis to the choice of electrostatic correction, Fig.~\ref{figS13} compares the stability diagrams obtained using only FNV corrections, while Fig. \ref{figS14} presents the corresponding results obtained exclusively using eFNV corrections.

\clearpage

\subsection*{S8. Supercell-size convergence of defect-induced optical spectra}

\begin{figure*}[ht]
\includegraphics[width=0.6\linewidth]{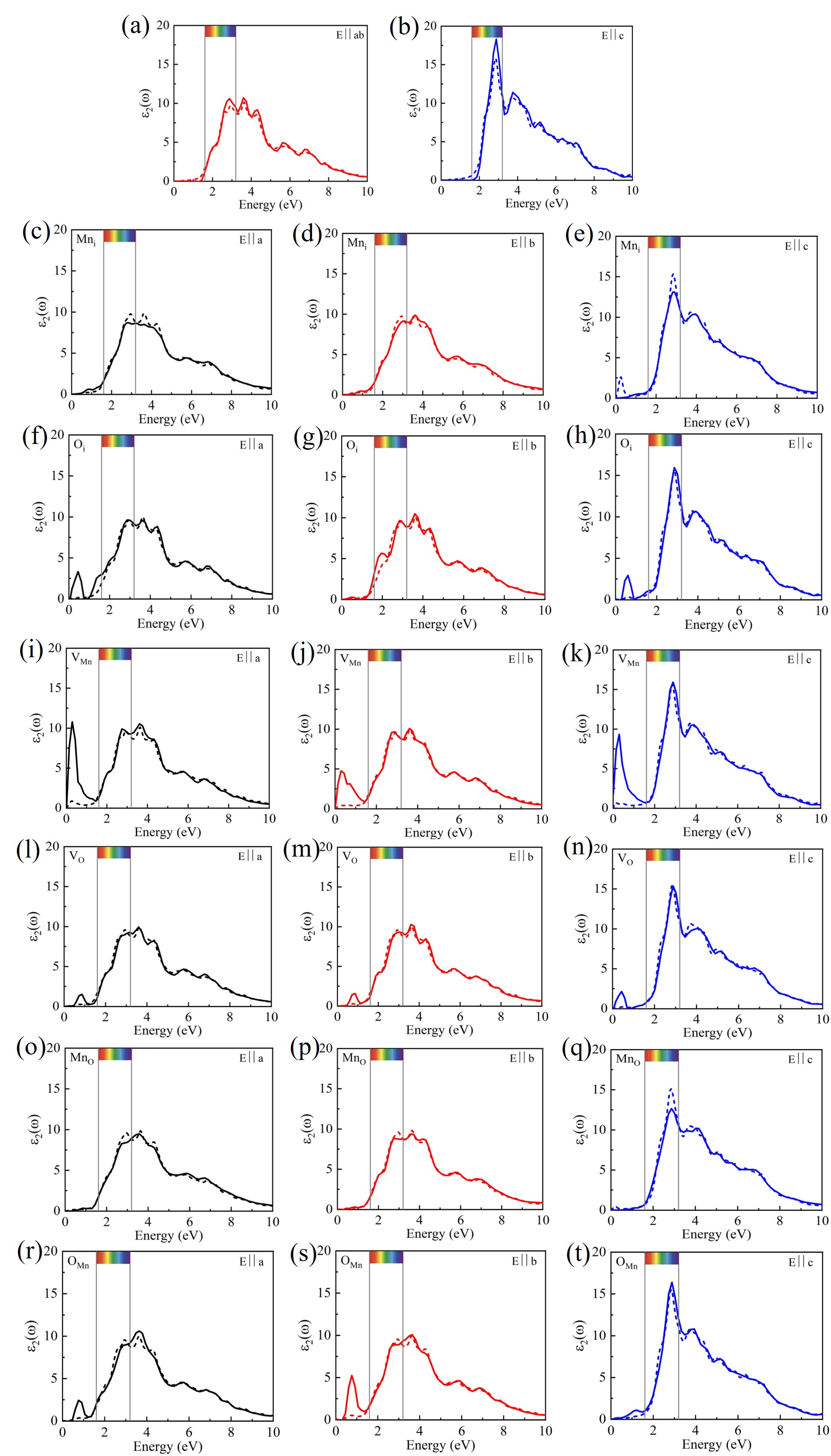}\caption{Imaginary part of dielectric constant $\varepsilon_2(\omega)$ for (a),(b) stoichiometric \(1 \times 1 \times 3\) (solid line) and \(2 \times 2 \times 7\) (dashed line) $\alpha$-MnO$_2$ supercell, and (c)-(t) isolated native defects, calculated within IPA.}
\label{figS15} 
\end{figure*}

Fig. \ref{figS15} compares the imaginary dielectric function of defective $\alpha$-MnO$_2$ calculated within the independent-particle approximation(IPA) using 1$\times$1$\times$3 and 2$\times$2$\times$7 supercells. The close agreement between the spectra confirms that the defect-induced features are not artifacts of the smaller supercell employed in the GW+BSE calculations. Although minor quantitative differences are present, the principal spectral features and the qualitative defect-dependent trends remain unchanged.

